\documentclass[a4paper,11pt]{article}
\usepackage{jheppub} 

\usepackage[T1]{fontenc}
\usepackage{braket, xcolor, soul, cleveref}
\let\vec\mathbf

\newcommand{\be}{\begin{equation}}
\newcommand{\ee}{\end{equation}}
\newcommand{\bea}{\begin{equation}\begin{aligned}}
\newcommand{\eea}{\end{aligned}\end{equation}}
\newcommand{\es}[2] {\begin{equation} \label{#1} \begin{split} #2 \end{split} \end{equation}}
\newcommand{\D}{\mathrm{d}}

\title{An Effective Cosmological Collider}

\author[a,b]{Nathaniel Craig,}
\author[c]{Soubhik Kumar,}
\author[b,d,e]{Amara McCune}

\affiliation[a]{Kavli Institute for Theoretical Physics, Santa Barbara, CA 93106, USA}
\affiliation[b]{Department of Physics, University of California, Santa Barbara, CA 93106, USA}
\affiliation[c]{Center for Cosmology and Particle Physics, Department of Physics,
New York University, New York, NY 10003, USA}
\affiliation[d]{Berkeley Center for Theoretical Physics, Department of Physics,
University of California, Berkeley, CA 94720, USA}
\affiliation[e]{Theoretical Physics Group, Lawrence Berkeley National Laboratory, Berkeley, CA 94720, USA}
\emailAdd{ncraig@physics.ucsb.edu}
\emailAdd{soubhik.kumar@nyu.edu}
\emailAdd{amara@physics.ucsb.edu}

\abstract{
Effective field theories (EFTs) of heavy particles coupled to the inflaton are rife with operator redundancies, frequently obscured by sensitivity to both boundary terms and field redefinitions. We initiate a systematic study of these redundancies by establishing a minimal operator basis for an archetypal example, the abelian gauge-Higgs-inflaton EFT. Working up to dimension 9, we show that certain low-dimensional operators are entirely redundant and identify new non-redundant operators with potentially interesting cosmological collider signals. Our methods generalize straightforwardly to other EFTs of heavy particles coupled to the inflaton.
}

\begin{document}
\setcounter{tocdepth}{2}
\maketitle
\flushbottom

\section{Introduction}
Local effective field theory (EFT) provides a powerful way to parametrize the effects of ultraviolet physics in a model-independent way. It has been employed with great success across a wide range of physical settings, from chiral perturbation theory and heavy quark effective theory to the EFT of inflation and the Standard Model (SM) EFT (see \cite{Baumgart:2022yty} and references therein for recent reviews). However, the simple recipe for constructing these EFTs --- by assembling the set of local, higher-dimensional operators consistent with the symmetries and field content of the infrared theory --- is often complicated by a vast redundancy of description associated with the insensitivity of observables to parameterizations of the fields.

Such redundancies are typically accommodated by identifying a `basis' of operators that is minimal, independent, and non-redundant with respect to the observables of interest. Operators outside the chosen basis can be expressed in terms of operators within the basis (or in some cases be eliminated entirely) in a variety of ways. For example, integration-by-parts (IBP) can be used to write certain operators in terms of others, dropping the boundary terms provided that fields vanish at spacetime infinity. Operators that differ by the lowest-order classical equations of motion (EOM) can be exchanged at linear order when on-shell quantities are concerned~\cite{Politzer:1980me, Georgi:1991ch, Arzt:1993gz, Henning:2017fpj}. In flat space, the LSZ reduction formula renders $S$-matrix elements insensitive to general field redefinitions. Reducing the scope of allowed EFT operators to a minimal basis with the methods at hand greatly facilitates the calculation of observables.

The simplifications arising from a minimal operator basis are perhaps most apparent when computing $S$-matrix elements in flat space, where EOM and IBP relations lead to a dramatic reduction in the number of operators at a given order in power counting \cite{Henning:2017fpj}. However, important differences arise in a cosmological context, particularly for an inflationary, quasi-de Sitter (dS) spacetime.
In this case, we are interested not in the $S$-matrix, but rather in the correlation functions of density perturbations at the end of inflation.
As a result, arbitrary field redefinitions are not allowed (or must be undone at some stage of the calculation) since such redefinitions would change the correlation functions.
Consequently, only a limited set of transformations can be used to remove redundant couplings. Furthermore, the correlation functions are computed on a late-time spatial boundary. As a result, temporal boundary terms may be relevant when implementing IBP relations.

These subtleties become particularly relevant in the context of primordial non-Gaussianity (NG).
The approximate scale invariance of primordial density perturbations, inferred through the cosmic microwave background (CMB), indicate that the interactions of the inflaton preserve an approximate shift symmetry, $\phi\rightarrow \phi+{\rm constant}$.
This implies both the self-interactions of the inflaton and interactions of the inflaton with other fields would involve operators with dimension 5 or higher, necessitating an EFT expansion.
The question of how to construct a minimal operator basis by eliminating redundant operators arises immediately. Unsurprisingly, there is a long history of identifying non-redundant self-interactions in the various EFTs of inflation \cite{Cheung:2007st, Weinberg:2008hq}, e.g.,~\cite{Maldacena:2002vr, Seery:2006tq, Arroja:2011yj, Rigopoulos:2011eq, Renaux-Petel:2011zgy, Ribeiro:2011ax,  deRham:2014wfa,  Bordin:2017hal, Garcia-Saenz:2019njm,   Bordin:2020eui, Green:2020ebl, Pajer:2020wxk, Ghosh:2023agt, Li:2023wdz}. 

In this work, we are interested in inflationary EFTs where additional heavy degrees of freedom are present.
Such EFTs are especially relevant for the `Cosmological Collider Physics' program~\cite{Chen:2009zp, Arkani-Hamed:2015bza} which aims to study oscillatory NG induced by {\it on-shell} particle production during inflation.
Particles with masses of order or larger than the inflationary Hubble scale $H_{\rm inf}$ can be produced as the Universe inflates.
Following their production, the heavy particles can oscillate in time, eventually decaying into inflaton fluctuations.
Those real-time oscillations of the heavy particles give rise to an oscillatory, scale-dependent NG.
Intriguingly, from the frequency of the oscillations and the angular dependence of the NG, one can extract the mass and spin of the heavy particle, respectively.
Since $H_{\rm inf}\lesssim 5\times 10^{13}$~GeV~\cite{Planck:2018jri}, the prospect of doing on-shell mass-spin spectroscopy of such heavy particles through NG provides a unique opportunity to study fundamental physics at high energies.
For various interesting work on this subject see Refs.~\cite{Baumann:2011su, Baumann:2011nk, Chen:2012ge, Noumi:2012vr, Assassi:2013gxa, Craig:2014rta, Dimastrogiovanni:2015pla, Meerburg:2016zdz, Lee:2016vti, Chen:2016uwp, Chen:2016nrs, Chen:2016hrz, An:2017hlx, Chen:2017ryl, Kumar:2017ecc, Baumann:2017jvh, MoradinezhadDizgah:2018ssw, Chen:2018xck, Arkani-Hamed:2018kmz, Kumar:2018jxz, Wu:2018lmx, Sleight:2019hfp, Lu:2019tjj, Hook:2019zxa, Hook:2019vcn, Kumar:2019ebj, Wang:2019gbi, Baumann:2019oyu, Li:2019ves, Alexander:2019vtb, Kogai:2020vzz, Bodas:2020yho, Aoki:2020zbj, Lu:2021gso, Lu:2021wxu, Wang:2021qez, Tong:2021wai, Cui:2021iie, Pimentel:2022fsc, Chen:2022vzh, AnilKumar:2022flx, Jazayeri:2022kjy, Qin:2022fbv, Niu:2022quw, Chen:2023txq, Qin:2023bjk, Aoki:2023dsl}.

Among the different types of theories that can give rise to a cosmological collider signal, gauge theories are particularly interesting.
Of course, such theories play a central role in the SM, and (hidden) gauge theories are also ubiquitous in physics beyond the SM.
With this motivation, in the present work we focus on the EFT of a gauge-Higgs sector coupled to the inflaton \cite{Kumar:2017ecc}, where the presence of a Higgs field allows us to incorporate spontaneous symmetry breaking (SSB) and study NG mediated by gauge bosons.\footnote{We clarify that by `Higgs' we will mean a generic complex scalar field that is not necessarily the SM Higgs. It can however be the SM Higgs, for example, if the electroweak scale is uplifted to $H_{\rm inf}$~\cite{Kumar:2017ecc}, or if loop corrections to SM are considered~\cite{Chen:2016hrz}.}
Furthermore, we focus on a $U(1)$ gauge theory, as our primary goal will be to lay out the procedure of operator basis construction in dS spacetime where heavy fields are present.
This can then be generalized to include non-Abelian gauge theories, which exhibit various interesting phenomena during inflation, such as thermalization and dissipation, as well as fermionic degrees of freedom, e.g.,~\cite{Anber:2009ua, Maleknejad:2011sq, Maleknejad:2012fw, Adshead:2012kp,Adshead:2018oaa,Domcke:2018eki}.

Establishing a minimal basis for the irrelevant interactions between the inflaton and an additional sector is essential to the determination of observable effects.\footnote{Of course, there are also more direct routes to mapping the space of cosmological observables that are free of EFT operator redundancies, as in the cosmological bootstrap (see, e.g.,~\cite{Baumann:2022jpr} for a recent review). Nonetheless, exploring observables from the standpoint of EFT Lagrangians can be useful for interpreting the microscopic implications of data and estimating the observability of certain signatures, motivating the approach taken here.} In doing so, one encounters subtleties analogous to those that arise in the study of inflationary self-interactions. Such a minimal basis relevant to cosmological collider signals was first developed in Ref.~\cite{Assassi:2013gxa} for a heavy real scalar field coupled to the inflaton at lowest non-trivial order. In this work, we perform a systematic construction of a minimal operator basis for the more general gauge-Higgs-inflaton EFT by considering operators up to dimension 9.
We impose an exact shift symmetry on the inflaton (discarding slow roll-suppressed corrections) and consider operators that describe the interactions of the inflaton with the gauge and the Higgs boson. To remove redundant operators, we primarily employ EOM and IBP relations. The EOM relations are closely analogous to those used in flat space \cite{Weinberg:2008hq}. 
On the other hand, IBP relations at dimension 5 do give rise to non-zero boundary terms that are {\it a priori} relevant. However, we show that such terms do not contribute to cosmological correlation functions involving heavy particles, and can be removed via appropriate field redefinitions. Beyond dimension 5, we find that IBP can indeed be used as in flat space for the operators of interest. One way to think about the irrelevance of these boundary terms is that they arise from interactions of the inflaton with heavy fields. The mode functions of the heavy fields decay at late times, corresponding to the physical effect that heavy particles get diluted as the Universe expands. Consequently, the temporal boundary terms resulting from IBP relations vanish for the operators and observables of interest.

To highlight some of our findings, we show that the only dimension 5 operator that is not redundant is an axionic coupling of the inflaton to the gauge field, namely $\phi F \tilde{F}$.
In particular, by using EOM we find that an operator coupling the Higgs current to the inflaton is redundant, and the leading Higgs-inflaton coupling arises only at dimension 6.
The same conclusion can be drawn using a field redefinition argument, as we discuss in an appendix.
We also show that in the broken phase of the theory, a quadratic mixing between the inflaton and the longitudinal mode of the gauge boson, relevant for tree-level bispectrum signatures, first arises at dimension 9.
Finally, we identify new operators at dimensions 7 and 8 involving the inflaton and gauge boson.
These operators are expected to contribute with similar strengths for NG compared to some other operators that have been considered in the previous literature.
While certain higher dimensional operators of the gauge-Higgs-inflaton theory have been considered in isolation in the previous literature, our systematic approach to enumerating an operator basis identifies additional operators that would be present in a generic EFT and could contribute to NG signals.

The rest of the article is organized as follows:
We discuss aspects of the choice of operator basis on inflationary observables in Sec.~\ref{sec:ibp}, with a particular emphasis on the effect of boundary terms generated by integration-by-parts manipulations.
In Sec.~\ref{sec:basis} we introduce an EFT of the inflaton coupled to an abelian gauge-Higgs theory, enumerating operators up to dimension 9 and reducing them to a minimal operator basis. Although boundary terms appear in certain cases, they do not affect the correlation functions of interest.
With the minimal basis at hand, we enumerate interaction vertices and estimate the leading sizes of non-Gaussianity in Sec.~\ref{sec:obs}.
We conclude in Sec.~\ref{sec:conc}.
A number of general considerations and examples regarding IBP and boundary terms in inflationary spacetimes can be found in the appendices.

\paragraph{Notations and Conventions.}
We follow the `mostly plus' metric signature: $(-,+,+,+)$.
The operator $\nabla_\mu$ denotes an ordinary covariant derivative, while the operator $D_\mu\equiv \nabla_\mu + i g_A A_\mu$ denotes a gauge covariant derivative. 
Unless explicitly stated, we will use units in which the Hubble scale during inflation, $H_{\rm inf} = 1$.
Factors of $H_{\rm inf} $ can be restored using dimensional analysis.

\section{Inflationary Observables and Operator Bases}\label{sec:ibp}

The precise nature of cosmic inflation is still unknown. Different classes of mechanisms can explain the homogeneous cosmic expansion and the generation of primordial fluctuations during inflation. To capture certain model-independent features and signatures, it is therefore useful to construct an EFT consistent with the symmetries and the particle content of the theory. In this regard, there are two qualitatively different classes of EFTs relevant during inflation. The more UV-agnostic of the two treats the inflaton as a Goldstone boson arising from spontaneous breaking of time translation symmetry~\cite{Cheung:2007st}. In this EFT, Lorentz invariance is (spontaneously) broken and as a result, one can allow qualitatively new sets of higher dimensional operators, in addition to the ones that follow from requiring Lorentz invariance.

Another class of EFT~\cite{Weinberg:2008hq} is useful if we are to describe both the inflationary fluctuations {\it and} the homogeneous inflationary expansion, since the latter is not necessarily part of the Goldstone EFT~\cite{Cheung:2007st}.
The advantage of this second class of EFTs is that it could be valid up to a much higher energy scale and can more readily describe how reheating can happen at the end of inflation. While the Lorentz-breaking, Goldstone EFT would allow for the greatest generality, for concreteness, in this work we will focus on a Lorentz-invariant EFT description and assume that inflation is driven by a slowly rolling scalar field $\phi$. Taking a bottom-up approach, we will also impose a strict shift symmetry on $\phi$: $\phi \rightarrow \phi+{\rm constant}$, motivated by the approximate scale invariance of primordial perturbations, and neglect subleading corrections from violation of this symmetry. Therefore, all the operators involving $\phi$ that we consider below will have (sometimes implicitly) derivative coupling $\nabla_\mu \phi$.

Our analysis will also encompass scenarios where the density fluctuations originate not from the inflaton $\phi$, but a curvaton field $\sigma$, as in the curvaton paradigm~\cite{Linde:1996gt, Enqvist:2001zp, Moroi:2001ct, Lyth:2001nq}.
In such scenarios, while the spacetime expansion is driven by $\phi$, the density fluctuations in the late Universe originate from $\sigma$. However, to obtain (approximately) scale-invariant, superhorizon fluctuations, the mass of $\sigma$ needs to be much smaller than $H_{\rm inf}$. Therefore, we can still impose a shift symmetry on $\sigma$.
Consequently, our following analysis will exactly carry over to the curvaton scenario, with the replacement $\nabla_\mu\phi \rightarrow \nabla_\mu \sigma$. With this in mind, in the rest of the discussion we will focus on the standard inflationary slow-roll EFT where both the homogeneous expansion and fluctuations are sourced by $\phi$.

\subsection{Minimal Operator Bases}

The approximate shift symmetry acting on the inflaton implies that the couplings between the inflaton and additional fields are necessarily irrelevant operators that may be organized systematically according to the relevant power-counting scheme. However, the full set of irrelevant operators consistent with the symmetries of the EFT is generally over-complete, leading to a redundancy of description whose severity depends on the observables of interest. In flat space where the observables are typically related to $S$-matrix elements, the redundancy of description corresponds to the freedom to perform nearly-arbitrary field redefinitions without altering the $S$-matrix elements. This can be used to arrive at minimal, non-redundant operator bases order-by-order in the power counting, where the number of operators in a minimal basis is typically much smaller than the total number of operators consistent with symmetries. In practice, a non-redundant basis of operators can usually be obtained order-by-order in power counting by using the lowest-order equations of motion to eliminate operators.\footnote{There are various subtleties involved when using equations of motion to eliminate operators, enumerated in \cite{Criado:2018sdb}.} Operators that differ by total derivatives can also be exchanged via integration-by-parts (IBP), as both spatial and temporal boundary terms are assumed to vanish in flat space.

The situation is somewhat different in cosmological contexts. In inflationary scenarios, we are interested in computing cosmological correlation functions at a fixed time slice towards the end of inflation, or when all the modes associated with the correlation function have exited the horizon. To be specific, we use the Poincare patch representation of dS spacetime
\es{}{
\D s^2 = {-\D \eta^2+ \D \vec{x}^2 \over \eta^2},
}
and denote the conformal time on the time slice of interest by $\eta=\eta_0$.
We then take $\eta_0\rightarrow 0$ limit of the cosmological correlators to obtain the conserved correlation functions.
Compared to the usual Minkowski spacetime, the presence of this boundary at $\eta_0$, where we evaluate the correlation functions, requires a reexamination of the standard operator basis manipulations, in particular those involving IBP.
This is because the boundary terms on the space-like surface at $\eta_0$ may not vanish in general.
The potential importance of boundary terms is already apparent for a massless free field in dS~\cite{Maldacena:2002vr}.

\subsection{Massless Free Field in dS}
Consider the Lagrangian of a free massless field and an equivalent expression obtained via IBP,
\es{}{
-{1\over 2}\int \D^4x \sqrt{-g}g^{\mu\nu}\nabla_\mu \phi \nabla_\nu \phi = -{1\over 2}\int \D^4x \sqrt{-g}g^{\mu\nu}\nabla_\mu (\phi \nabla_\nu \phi)+{1\over 2}\int \D^4x \sqrt{-g}\phi \square \phi,
}
where $\square \phi = g^{\mu\nu}\nabla_\mu\nabla_\nu\phi$.
Using the EOM $\square\phi\approx 0$ (neglecting $V(\phi)$) and Stokes' theorem we arrive at
\es{}{
-{1\over 2}\int \D^4x \sqrt{-g}g^{\mu\nu}\nabla_\mu \phi \nabla_\nu \phi = -{1\over 2}\int \D^3x \sqrt{\gamma}n_\mu (\phi \nabla^\mu \phi).
}
Here the spatial integration is over a space-like surface at $\eta_0$ and we have assumed the fields vanish at spatial infinity as well as at very early times when the fields are in their vacuum states.
The vector $n_\mu = (1/\eta_0, 0,0,0)$ is normal to the space-like surface on which the induced metric is given by $\gamma_{ij}$, with a determinant $\gamma = 1/\eta_0^6$.
Thus the (on-shell) action of a massless free field in dS can be written as a boundary term at $\eta_0$ which can be simplified as,
\es{eq:inf_bd}{
{1\over 2\eta_0^2}\int \D^3x  \left.\phi \partial_\eta \phi\right\rvert_{\eta_0} = {1\over 2\eta_0^2}\int {\D^3k\over (2\pi)^3}\left.\phi_{\vec{k}}\partial_\eta \phi_{-\vec{k}}\right\rvert_{\eta_0}.
}
To compute correlation functions, we can first derive the associated wavefunction which can be schematically written as (see~\cite{Baumann:2022jpr} for a recent review),
\es{}{
\Psi[\varphi,\eta_0] = \int_{\phi(\eta_0)=\varphi,\phi(-\infty)=0} {\cal D}\phi e^{i S[\phi]} \propto e^{iS_{\rm cl}[\varphi,\eta_0]},
}
where $\phi(\eta_0)=\varphi$ is the late time boundary condition while $\phi(-\infty)=0$ ensures that the fields are in their vacuum state at early times.
We have also evaluated the path integral using the saddle point approximation, up to a proportionality constant.
With these choices, we can write
\es{eq:phi_late}{
\phi_\vec{k} = \varphi_\vec{k}{(1-ik\eta)e^{ik\eta} \over (1-ik\eta_0)e^{ik\eta_0}}.
} 
The classical action is then given by (upon using $\varphi_{-\vec{k}} = \varphi_{\vec{k}}^*$),
\es{eq:freeScl}{
S_{\rm cl} = {1\over 2\eta_0^2}\int {\D^3k\over (2\pi)^3}{k^2\eta_0 \over (1-ik\eta_0)} |\varphi_{\vec{k}}|^2.
}
A correlation function at a late time, $\eta_0$ is given by
\es{}{
\langle \varphi(\vec{k}_1) \varphi(\vec{k}_2)\cdots \varphi(\vec{k}_n)\rangle = {\int {\cal D}\varphi \varphi(\vec{k}_1) \varphi(\vec{k}_2)\cdots \varphi(\vec{k}_n) |\Psi[\varphi,\eta_0]|^2 \over \int {\cal D}\varphi |\Psi[\varphi,\eta_0]|^2}.
}
Given the appearance of $|\Psi[\varphi,\eta_0]|^2$, only the imaginary terms in $S_{\rm cl}$ contribute to the determination of the correlation function.
From~\eqref{eq:freeScl}, the part that diverges as $\eta_0\rightarrow 0$ then does not contribute and the surviving contribution is given by
\es{}{
S_{\rm cl} \approx \int {\D^3 k \over (2\pi)^3} {ik^3 \over 2} |\varphi_{\vec{k}}|^2. 
}
We can now evaluate the two-point function using this wavefunction,
\es{}{
\langle \varphi(\vec{k}_1) \varphi(\vec{k}_2)\rangle &= {\int {\cal D}\varphi \varphi(\vec{k}_1) \varphi(\vec{k}_2) |\Psi[\varphi,\eta_0]|^2 \over \int {\cal D}\varphi |\Psi[\varphi,\eta_0]|^2},\\
&= {\int {\cal D}\varphi \varphi(\vec{k}_1) \varphi(\vec{k}_2) \exp\left(-\int {\D^3 k \over (2\pi)^3} {k^3 \over 2} |\varphi_{\vec{k}}|^2\right) \over \int {\cal D}\varphi \exp\left(-\int {\D^3 k \over (2\pi)^3} {k^3 \over 2} |\varphi_{\vec{k}}|^2\right)},\\
&= {1 \over 2k_1^3}(2\pi)^3\delta(\vec{k}_1+\vec{k}_2),
}
as can also be derived using the standard `in-in' computation (see \cite{Chen:2017ryl} for a pedagogical review).
This example illustrates that the imaginary part of $S_{\rm cl}$ is relevant for computing the correlation function, while the real part drops out from $|\Psi[\varphi,\eta_0]|^2$.

There is another way to reach the same conclusions as above, highlighting the role of the boundary.
We can treat the fields in~\eqref{eq:inf_bd} as quantum operators, instead of classical functions.
The two-point function can then be computed using the standard `in-in' approach.
Since the correlation functions are evaluated on the spatial surface and Eq.~\eqref{eq:inf_bd} is also evaluated on the same surface, we need to evaluate only equal-time propagators.
We also note that the usual time evolution operator $T(\exp(-i \int \D t~{\mathbb H}))$, with $\mathbb{H}$ the Hamiltonian, can be schematically written as a boundary term $\sim \exp(-i f(\eta_0))$.
Thus the time ordering part gives a factor of $(-i)$, while the anti-time ordering part gives the complex conjugate factor $(+i)$, as in a standard bulk computation.
With these ingredients, the result is given by
\es{}{
\langle\phi(\vec{k}_1)\phi(\vec{k}_2)\rangle' & = 2\times (-i){1\over 2\eta_0^2}{1 \over 4k_1^6}(1+ik_1\eta_0)^2(1-ik_1\eta_0)k_1^2\eta_0 +{\rm c.c.}\\
& = {1\over \eta_0}{1 \over 4k_1^4}(-i+k_1\eta_0)(1+k_1^2\eta_0^2) +{\rm c.c.}\\
& = {1\over 2k_1^3} + {\cal O}(\eta_0).
}
Here we have used the standard notation $\langle\phi(\vec{k}_1)\phi(\vec{k}_2)\rangle \equiv (2\pi)^3 \delta(\vec{k}_1+\vec{k}_2) \langle\phi(\vec{k}_1)\phi(\vec{k}_2)\rangle'$. Further examples of the relevance of boundary terms for massless and massive scalars, both free and with derivatively-coupled cubic interactions, are presented in Appendix \ref{app:ex}.

\subsection{Operators Coupled to the Inflaton}
In what follows, we will use arguments similar to those presented above to understand whether certain boundary terms contribute or not in determining cosmological correlators. While the boundary term was essential in the previous example, in many cases it can be neglected. In particular, we often encounter operators of the type
\es{}{
\int \D^4 x \sqrt{|g|} \nabla_\mu \phi \nabla^\mu {\cal O},
}
where ${\cal O}$ is any (composite) operator containing massive fields.
Using an IBP we can write the above as
\es{}{
\int \D^4 x \, \sqrt{|g|} \left( \nabla^\mu\left[ \nabla_\mu \phi \cdot {\cal O}\right] - \square\phi\cdot {\cal O}\right).
}
The second term does not contribute in any vertex for an `in-in' diagram since $\square f = 0$ where $f$ is a mode function for the inflaton.
The first term, on the other hand, can be written as
\es{}{
\int \D^4 x \, \partial_\mu \left(\sqrt{|g|}\nabla^\mu\phi\cdot {\cal O} \right).
}
For the spatial component, i.e., for $\mu=i$, the above does not contribute under the assumption that fields vanish at spatial infinity.
Therefore, the only potentially non-trivial term is the one involving time derivatives,
\es{}{
-\int \D^4 x \, \partial_\eta\left({1\over \eta^2}\partial_\eta\phi\cdot {\cal O}\right).
}
This determines the interacting Hamiltonian of interest,
\es{}{
\mathbb{H}_I = \int \D^3 x \, \partial_\eta \left({1\over \eta^2}\partial_\eta\phi\cdot {\cal O}\right).
}
However, since this is a total time derivative, we can evaluate the time evolution operator as
\es{}{
T \exp\left(-i\int_{-\infty}^{\eta_0}\D \eta \,{\mathbb{H}_I}\right) = T \exp\left(-i\int \D^3 x \left[{1\over \eta^2}\partial_\eta \phi \cdot {\cal O}\right]_{\eta_0}\right).
}
The last term is evaluated at $\eta_0$ and it does not involve any time integrals.
Therefore, the time ordering operator acts trivially.
We can then conclude that if the term in the square brackets vanishes at $\eta_0$, the entire operator does not contribute to correlation functions involving the massive particle.\footnote{A detailed example of the action of the time ordering operator can be found in Appendix~\ref{app.detailed_example}, where we show that while boundary terms can be present in the intermediate stages of a computation, they do not give a (non-local) cosmological collider signal with the characteristic non-analytic momentum dependence. Rather, the boundary terms give a local contribution which can be accounted for by appropriate local field redefinitions.}

\section{The Abelian Gauge-Higgs-Inflaton EFT}\label{sec:basis}

We're now equipped to construct minimal operator bases in dS for EFTs where a shift-symmetric inflaton couples to additional fields. For concreteness, in this article we will take the additional fields to comprise an abelian gauge-Higgs sector. This theory is of considerable interest as a benchmark for various cosmological collider signals, and captures most of the relevant features involved in constructing a minimal operator basis. A similar procedure can be followed for other effective theories containing different fields, such as fermions or non-abelian gauge bosons.

The Lagrangian containing the inflaton ($\phi$), a Higgs (${\cal H}$), and a $U(1)$ gauge field $A_\mu$, up to dimension-4 is given by
\es{eq:lag_dim_4}{
{\cal L} \supset -{1\over 2}\nabla_\mu\phi\nabla^\mu\phi-V(\phi)-(D_\mu {\cal H})^\dagger D^\mu {\cal H}-V(|{\cal H}|^2) - {1\over 4} F_{\mu\nu}F^{\mu\nu}.
}
Here $D_\mu {\cal H} = \nabla_\mu {\cal H} + i g_A A_\mu {\cal H}$ is the gauge covariant derivative and $V(|{\cal H}|^2)$, $V(\phi)$ are respectively the Higgs and inflaton potentials, whose detailed forms will not be important for our purposes. We will see in some cases that the surviving contributions from a given operator are slow roll-suppressed, in the sense of involving $\D V(\phi)/\D \phi$ or $\D^2V(\phi)/\D \phi^2$. We will not track such operators explicitly, under the assumption that their slow roll-suppressed contributions to observables are subdominant to other contributions. We assume the Higgs potential is such that it can acquire a vacuum expectation value $\left \langle {\cal H} \right \rangle = v/\sqrt{2}$, so that all states in the abelian gauge-Higgs sector are massive in the broken phase. Alternately, our results can also be applied to the theory of a shift-symmetric inflaton coupled to a complex scalar with a global $U(1)$ symmetry by taking the $g_A \rightarrow 0$ limit and assuming $\left \langle {\cal H} \right \rangle = 0$.

In reducing the operator basis, we will primarily use the following EOM and also implement IBP.
The EOM for the inflaton is given by,
\es{eq:eom_inf}{
\square\phi=V'(\phi),~~~{\rm [Inflaton~EOM]}
}where we have denoted $\square \equiv \nabla^\mu\nabla_\mu$.
To obtain the EOM for the Higgs, we first expand
\es{}{
-(D_\mu {\cal H})^\dagger D^\mu {\cal H} = -\nabla_\mu {\cal H}^\dagger \nabla^\mu {\cal H} + i g_A A_\mu {\cal H}^\dagger \nabla^\mu {\cal H} - ig_A A^\mu {\cal H}\nabla_\mu {\cal H}^\dagger - g_A^2 A_\mu A^\mu {\cal H}^\dagger {\cal H}.
}
The EOM is then given by,
\es{eq:eom_H}{
-\nabla_\mu\left(\nabla^\mu {\cal H}+ig_A A^\mu {\cal H}\right) =+ig_A A_\mu \nabla^\mu {\cal H} - g_A^2 A_\mu A^\mu {\cal H} - V'(|{\cal H}|^2){\cal H},
}
which can be written in terms of the gauge covariant derivative,
\es{eq:eom_H}{
-\nabla_\mu D^\mu {\cal H} = ig_A A_\mu D^\mu {\cal H} - V'(|{\cal H}|^2){\cal H}.~~~{\rm [Higgs~EOM]}
}
The EOM for the gauge field is given by
\es{eq:eom_A}{
-\nabla_\mu F^{\mu\nu} = ig_A\left({\cal H}^\dagger D^\nu {\cal H}-(D^\nu {\cal H})^\dagger {\cal H}\right).~~~{\rm [Gauge~Field~EOM]}
}

The symmetries of the theory forbid relevant or marginal couplings between the inflaton and the abelian gauge-Higgs sector, so interactions are necessarily irrelevant. We assume the gauge-Higgs sector is weakly coupled and the Higgs VEV is parametrically smaller than the characteristic UV scale $\Lambda$ suppressing the irrelevant operators, so that the natural power counting is in terms of the classical dimension of operators constructed out of the fields in the unbroken phase. 

In what follows, we enumerate operators up to dimension 9, beginning with the complete set of operators at a given dimension allowed by symmetries, modulo some operators trivially related by EOM. We then reduce the operators to a minimal basis at a given order via EOM and IBP relations, taking care to verify that boundary terms arising from IBP do not contribute to the observables of interest. Since we are taking a bottom-up approach in which the Wilson coefficients of different operators are free parameters, we may use the lowest-order EOM to arrive at a minimal basis at a given order in power-counting. Although these EOM manipulations do not correctly account for changes to Wilson coefficients at higher order in power counting, the values of these coefficients were already arbitrary. In this way, we can fix the operator basis using the lowest-order EOM by starting at dimension-5 and proceeding to successively higher dimensions. Note that more care would be required in manipulating operator bases when matching to a specific UV completion in which all Wilson coefficients take on specific values \cite{Criado:2018sdb}. 

Needless to say, the number of possible operators grows rapidly with operator dimension. Although it is not too cumbersome to enumerate operators up to dimension-9 by hand, we have also cross-checked our results against the flat-space operator basis codes {\tt DEFT} \cite{Gripaios:2018zrz} and {\tt Sym2Int} \cite{Fonseca:2017lem, Fonseca:2019yya}.

\subsection{Dimension 5}
We start our analysis of irrelevant operators at dimension 5. While there are a number of operators consistent with the assumed symmetries, we show there is only one operator that contributes non-trivially to cosmological correlators.
We first summarize the operators in Table~\ref{tab:dim5}. The Wilson coefficient for each operator is taken to be real; note that operators such as $\mathcal{O}_{5,1}$ and $\mathcal{O}_{5,2}$ can be interpreted as the real and imaginary parts of a single operator with a complex Wilson coefficient. 
\begin{table}
	\begin{center}
  \begin{tabular}{ | c | c |}
    \hline
    Operator & Expression \\ \hline
    \hline
    ${\cal O}_{5,1}$ & $\nabla_\mu\phi \left({\cal H}^\dagger D^\mu {\cal H} + (D^\mu {\cal H})^\dagger {\cal H}\right)$ \\ 
    \hline
    ${\cal O}_{5,2}$ & $(-i)\nabla_\mu\phi \left({\cal H}^\dagger D^\mu {\cal H} - (D^\mu {\cal H})^\dagger {\cal H}\right)$ \\
    \hline
    ${\cal O}_{5,3}$ & $\nabla_\mu\phi \nabla_\nu F^{\nu\mu}$ \\
    \hline
    ${\cal O}_{5,4}$ & $\phi F_{\mu\nu} \tilde{F}^{\mu\nu}$\\
    \hline
  \end{tabular}
\end{center}
\caption{Allowed operators at dimension 5. Here and henceforth, $ \tilde{F}^{\mu\nu} = (1/2)\epsilon^{\mu\nu\rho\sigma}F_{\rho\sigma}.$}
\label{tab:dim5}
\end{table}

We note that ${\cal O}_{5,1}$ can be simplified as
\es{}{
{\cal O}_{5,1} = \nabla_\mu \phi \nabla^\mu({\cal H}^\dagger {\cal H}).
}
We can use the EOM~\eqref{eq:eom_A} for the gauge field and subsequently an IBP to write
\es{}{
{\cal O}_{5,2} = {1\over g_A}\nabla_\nu\left(F^{\nu\mu} \nabla_\mu\phi\right).
}
In the process we have dropped a contribution of the type $\nabla_\nu\nabla_\mu\phi\cdot F^{\mu\nu}$ which vanishes identically for a torsion-free metric.
This manipulation also shows that ${\cal O}_{5,2}$ is equivalent to ${\cal O}_{5,3}$.
To comprehensively study the fate of ${\cal O}_{5,1}$ and ${\cal O}_{5,2}$, it is useful to consider both the unbroken and broken phases of the theory.

\subsubsection{Unbroken Phase}
We start with ${\cal O}_{5,1}$ which after an IBP gives,
\es{}{
{\cal O}_{5,1}= \int \D^4 x \sqrt{-g} \left[ \nabla^\mu(\nabla_\mu\phi\cdot {\cal H}^\dagger {\cal H})- \square\phi \cdot {\cal H}^\dagger {\cal H}\right].
}
The second term vanishes in the limit of vanishing inflaton potential and we will not consider it further.
The first term is a boundary term that reduces to a spatial integral at $\eta_0$,
\es{}{
{\cal O}_{5,1}=\int \D^3 x\sqrt{\gamma} n^\mu (\nabla_\mu\phi\cdot {\cal H}^\dagger {\cal H}) = -{1 \over \eta_0^2}\int \D^3x \partial_\eta \phi\cdot {\cal H}^\dagger {\cal H}.
}
From~\eqref{eq:phi_late}, we note
\es{}{
\partial_\eta \phi_{\vec{k}} = \varphi_{\vec{k}} {k^2\eta_0 \over (1-ik\eta_0)}.
}
Given the late time scaling of ${\cal H}(\eta, \vec{k}) \sim \eta_0^{3/2\pm i\mu}$ (as can be shown by considering mode functions of a massive particle, see, e.g.,~\cite{Chen:2022vzh}), where $\mu \equiv \sqrt{m^2/H_{\rm inf}^2-9/4}$ is taken to be positive, we have
\es{}{
{\cal O}_{5,1} \propto \eta_0^2 \rightarrow 0.~~{\rm [Unbroken~Theory]}
}
Therefore, ${\cal O}_{5,1}$ does not contribute to late-time correlators. 

We now consider ${\cal O}_{5,2}$ which is also a boundary term and can be rewritten as,
\es{}{
{\cal O}_{5,2} = {1\over g_A}\int \D^3 x \sqrt{\gamma} n_\nu \nabla_\mu\phi F^{\nu\mu} = -{1\over g_A}\int \D^3 x  \partial_i\phi F_{\eta i}.
}
For an unbroken gauge theory, the physical degrees of freedom are the transverse components $A_i^\perp$ which satisfy $k_i A_i^\perp = 0$.
Therefore, ${\cal O}_{5,2}$ does not contribute to late-time correlators.
Note, that this argument does not rely on the vanishing of the mode functions at late times.

\subsubsection{Broken Phase}
We now repeat the above analysis for the case of a broken gauge theory starting with ${\cal O}_{5,1}$.
We can still implement an IBP to write it as
\es{}{
{\cal O}_{5,1} = -{1 \over \eta_0^2}\int \D^3x \partial_\eta \phi\cdot {\cal H}^\dagger {\cal H}.
}
In the broken gauge theory, we can set one of the Higgs to its VEV to obtain a term quadratic in fluctuations.
However, the result still scales as,
\es{}{
{\cal O}_{5,1} \propto \eta_0^{1/2}\rightarrow 0~~{\rm [Broken~Theory]}.
}
Hence ${\cal O}_{5,1}$ does not contribute in the broken gauge theory as well.

Next we turn to ${\cal O}_{5,2}$, which can be written as
\es{eq:o52_simple}{
{\cal O}_{5,2} = -{1\over g_A}\int \D^3 x  \partial_i\phi F_{\eta i} = -{1\over g_A}\int \D^3 x  \partial_i\phi (\partial_\eta A_i^{\parallel} - \partial_i A_\eta).
}
However, we need to take into account the longitudinal component of the gauge boson which is a combination of $A_\eta$ and $A_i^{\parallel}$.
The temporal component $A_\eta$ falls as $\eta_0^{3/2\pm i\mu}$, with $\mu \equiv (m^2/H_{\rm inf}^2-1/4)^{1/2} >0$, at late times, as can be seen from the massive gauge boson mode functions, e.g.,~\cite{Lee:2016vti}.
Therefore the term proportional to $\partial_i A_\eta$ in~\eqref{eq:o52_simple} vanishes in the late time limit.
After a spatial IBP, the surviving term can be written as
\es{}{
{\cal O}_{5,2} = {1\over g_A}\int \D^3 x \phi \partial_\eta \partial_i A_i^{\parallel}.
}
We can rewrite the above after using the constraint equation for the massive field, 
\es{eq:AconEq}{
\nabla_\mu A^\mu = 0 \Rightarrow -\eta^2\partial_\eta A_\eta+2\eta A_\eta +\eta^2 \partial_i A_i^{\parallel}=0,
}
\es{}{
{\cal O}_{5,2} = {1\over g_A}\int \D^3 x \, \phi \partial_\eta \left(\partial_\eta A_\eta - {2\over \eta}A_\eta\right).
}
This can be further simplified by using the EOM for $A_\eta$,
\es{eq:AetaEOM}{
\partial_\eta^2 A_\eta - {2\over \eta}\partial_\eta A_\eta - \partial_i^2 A_\eta + {m^2+2 \over \eta^2}A_\eta = 0,
}
\es{}{
{\cal O}_{5,2} = -{m^2\over g_A \eta_0^2}\int \D^3 x \, \phi A_\eta.
}
While this term does not vanish as $\eta_0\rightarrow 0$, it does not contribute to a late-time correlation function.
As an example, we can evaluate the contribution to the two-point inflaton correlation function from ${\cal O}_{5,2}$.
That has a scaling:
\es{}{
\langle \phi(\vec{k}_1) \phi(\vec{k}_2)\rangle \propto {1\over \eta_0^4}\times \eta_0^3(1+k_1^2\eta_0^2) (1+k_1^2\eta_0^2)((-i)^2 + (+i)^2+(+i)(-i) + (+i)(-i)) =0,
}
where the last factor in the parenthesis comes from summing over the four `in-in' subdiagrams, while the factor of $\eta_0^3$ comes from the massive $A_\eta$ propagator at late times.

More generally, each factor of ${\cal O}_{5,2}$ appears in a correlator involving the inflaton field on the late time surface with a structure 
\es{}{
\propto (-i)(1-ik\eta_0) {|f_\eta(\eta_0,k)| \over \eta_0^2}+ (+i)(1+ik\eta_0) {|f_\eta(\eta_0,k)| \over \eta_0^2} = -2k {|f_\eta(\eta_0,k)| \over \eta_0},
}
where $A_\eta(\eta,\vec{k}) = f_\eta(\eta,k) b_{\vec{k}}^\dagger + f^*_\eta(\eta,k) b_{-\vec{k}}$, with $f_\eta(\eta,k)$ a mode function and $b_{\vec{k}}^\dagger$ a creation operator, along with their conjugates.
The prefactors are $(-i)$ or $(+i)$ depending on whether the contribution originally came from a time ordering or anti-time ordering term.\footnote{Since here we are interested in evaluating just surface terms at $\eta_0$, the action of time ordering or anti-time ordering is trivial. However, the factors of $(-i)$ or $(+i)$ are still present depending on whether terms originally came from time ordering operator $T \exp(-i \int \D t~{\mathbb H})$ or anti-time ordering operator $(T \exp(-i \int \D t~{\mathbb H}))^\dagger$.}
We have also kept only the absolute value $|f_\eta(\eta_0,k)|$ since, in an inflaton correlator on the late time surface, we always have the combination $|f_\eta(\eta_0,k)|^2$ from the longitudinal mode propagator, so each factor of $A_\eta$ effectively contributes a factor of $|f_\eta(\eta_0,k)|$.
Noting that $|f_\eta(\eta_0,k)| \sim \eta_0^{3/2}$ for $\mu>0$, we see ${\cal O}_{5,2}$ does not contribute to a correlation function from contractions on the late time surface.
It can also be checked that bulk contractions with the operators in Table~\ref{table:summary} vanish as $\eta_0\rightarrow 0$.
Therefore, ${\cal O}_{5,2}$ does not contribute to cosmological correlation functions overall.

\subsubsection{Surviving Contribution at Dimension 5}
The above analysis shows that the three operators ${\cal O}_{5,1}$, ${\cal O}_{5,2}$, and ${\cal O}_{5,3}$ are all redundant.
The remaining operator is ${\cal O}_{5,4}$ which is also shift symmetric since $F_{\mu\nu}\tilde{F}^{\mu\nu}$ is a total derivative.
This operator is non-trivial and has been discussed extensively in the context of gauge field production during inflation (see, e.g.,~\cite{Anber:2009ua}), and cosmological collider~\cite{Wang:2020ioa}.

\subsection{Dimension 6}
At dimension 6, we only have one operator coupling the inflaton field to the gauge-Higgs sector,
\begin{align}
    \mathcal{O}_{6,1} \equiv \left(\nabla_\mu \phi \right)^2 {\cal H}^\dagger {\cal H}.
\end{align}
This term is not reducible to another operator, and it contributes both in the broken and unbroken phase, via tree and loop-level diagrams, respectively.
The associated signatures have been discussed in~\cite{Chen:2016nrs, Kumar:2017ecc}.

\subsection{Dimension 7}
\label{sec:dim_7}
At this dimension, there are four possible operators to start with, as summarized in Table~\ref{tab:dim7}.
\begin{table}
	\begin{center}
  \begin{tabular}{ | c | c |}
    \hline
    Operator & Expression \\ \hline
    \hline
    ${\cal O}_{7,1}$ & $|{\cal H}|^2 \nabla_\mu \phi \left({\cal H}^\dagger D^\mu {\cal H} + (D^\mu {\cal H})^\dagger {\cal H}\right)$ \\ 
    \hline
    ${\cal O}_{7,2}$ & $|{\cal H}|^2 \nabla_\mu \phi \left({\cal H}^\dagger D^\mu {\cal H} - (D^\mu {\cal H})^\dagger {\cal H}\right)$ \\
    \hline
    ${\cal O}_{7,3}$ & $F^{\mu\nu} \nabla_\mu \phi  \left({\cal H}^\dagger D_\nu {\cal H} + (D_\nu {\cal H})^\dagger {\cal H}\right)$ \\
    \hline
    ${\cal O}_{7,4}$ & $(-i) F^{\mu\nu} \nabla_\mu \phi  \left({\cal H}^\dagger D_\nu {\cal H} - (D_\nu {\cal H})^\dagger {\cal H}\right)$\\
    \hline
  \end{tabular}
\end{center}
\caption{Allowed operators at dimension 7.}
\label{tab:dim7}
\end{table}
We start the analysis with ${\cal O}_{7,1}$ which after IBP can be written as
\es{}{
{\cal O}_{7,1} = \nabla_\mu \phi \cdot {\cal H}^\dagger {\cal H} \nabla^\mu ({\cal H}^\dagger {\cal H}) =  {1\over 2}\nabla_\mu \phi \nabla^\mu |{\cal H}|^4 ={1\over 2}\nabla^\mu \left( |{\cal H}|^4 \nabla_\mu \phi\right),
}
where in the last line we have dropped a contribution proportional to $\square \phi$ as that is slow roll-suppressed.
The surviving term is a boundary term, which we can evaluate by following the procedure detailed in the previous section.
On a late time surface at $\eta_0$, it scales as
\es{}{
{\cal O}_{7,1} \sim \sqrt{\gamma} n_\eta g^{\eta\eta} |{\cal H}|^4 \partial_\eta\phi \sim {1\over \eta_0^4} \times \eta_0^2 \times |{\cal H}|^4 \times \eta_0.
}
Now we need to set at least one of the Higgs to its fluctuation, otherwise we would just have a dimension 5 operator considered before.
Since the ${\cal H}$ mode function has a scaling $\eta_0^{3/2}$, up to oscillatory parts, we see that the entire boundary term vanishes.
Therefore ${\cal O}_{7,1}$ does not contribute.

Next we consider ${\cal O}_{7,2}$.
We can use the EOM for the gauge field to write
\es{}{
{\cal O}_{7,2} = {i\over g_A} |{\cal H}|^2 \nabla_\mu \phi \nabla_\nu F^{\nu\mu}.
}
However, this is equivalent to ${\cal O}_{7,3}$.
To see this, we write,
\es{}{
{\cal O}_{7,3} = F^{\mu\nu} \nabla_\mu\phi \nabla_\nu({\cal H}^\dagger {\cal H}) = \nabla_\nu \left(F^{\mu\nu} \nabla_\mu\phi ({\cal H}^\dagger {\cal H}) \right) - \nabla_\nu F^{\mu\nu}\cdot \nabla_\mu\phi \cdot {\cal H}^\dagger {\cal H}.
}
On the late-time surface, the total derivative term scales as
\es{}{
{1\over \eta_0^4} \times \eta_0^4 \partial_\eta A_i^{\parallel} |{\cal H}|^2.
}
We cannot set both the ${\cal H}$ to their VEVs, otherwise we just go back to a dimension 5 operator considered above. 
Therefore, since at least one factor of the ${\cal H}$ fluctuation has to be present, going as $\eta_0^{3/2}$, the entire term scales at least as $\eta_0$.
Thus the total derivative does not contribute.
Finally we consider ${\cal O}_{7,4}$, which upon using the gauge field EOM becomes
\es{}{
{\cal O}_{7,4} ={1 \over g_A} F_{\mu\nu}\nabla^\mu\phi \nabla_\rho F^{\rho \nu},
}
and can contribute to correlation functions. In total there are two non-redundant operators at dimension 7, neither of which have yet been considered in the literature.

\subsection{Dimension 8}
Many more operators are allowed at dimension 8, as enumerated in Table~\ref{tab:dim8}.
\begin{table}
	\begin{center}
  \begin{tabular}{ | c | c |}
    \hline
    Operator & Expression \\ \hline
    \hline
    $\mathcal{O}_{8,1}$ &  $F_{\mu \nu} F^{\mu \nu} (\nabla_\rho \phi)^2$ \\ 
    \hline
    $\mathcal{O}_{8,2}$ & $F_{\mu \nu} \tilde F^{\mu \nu} (\nabla_\rho \phi)^2$ \\
    \hline
    $\mathcal{O}_{8,3}$ &  $|{\cal H}|^4 (\nabla_\mu \phi)^2$\\
    \hline
    $\mathcal{O}_{8,4}$ & $|D_\mu {\cal H}|^2 (\nabla_\nu \phi)^2$\\
    \hline
    $\mathcal{O}_{8,5}$ & $(D^\mu {\cal H})^\dag D^\nu {\cal H} \nabla_\mu \phi \nabla_\nu \phi$\\
    \hline
    $\mathcal{O}_{8,6}$ & $F_{\mu \rho} F^{\nu \rho} \nabla^\mu \phi \nabla_\nu \phi$\\
    \hline
    $\mathcal{O}_{8,7}$ & $F_{\mu \rho} \tilde F^{\nu \rho} \nabla^\mu \phi \nabla_\nu \phi$ \\ 
    \hline
$\mathcal{O}_{8,8}$ & $\tilde F_{\mu \rho} \tilde F^{\nu \rho} \nabla^\mu \phi \nabla_\nu \phi$ \\
\hline
$\mathcal{O}_{8,9}$ & $({\cal H}^\dag D^\mu {\cal H} + (D^\mu {\cal H})^\dagger {\cal H}) \nabla_\mu \nabla_\nu \phi \nabla^\nu \phi$\\
\hline
$\mathcal{O}_{8,10}$ & $(-i)({\cal H}^\dag D^\mu {\cal H} - (D^\mu {\cal H})^\dagger {\cal H}) \nabla_\mu \nabla_\nu \phi \nabla^\nu \phi$\\
\hline
\end{tabular}
\end{center}
\caption{Allowed operators at dimension 8.}
\label{tab:dim8}
\end{table}
The operators $\mathcal{O}_{8,1} - \mathcal{O}_{8,6}$ form a non-redundant minimal basis. Of the remaining operators, ${\cal O}_{8,7}$ simply vanishes.
To see this, we fix $\mu$ and $\nu$ such that they are not identical.
If they are identical, then the associated contribution is already a part of ${\cal O}_{8,2}$.
Then we can write,
\es{}{
F_{\mu\rho}\tilde{F}^{\nu\rho} = {1\over 2}F_{\mu\rho}\epsilon^{\nu\rho\alpha\beta} F_{\alpha\beta}.
}
The index $\rho$ has to be different from both $\mu$ and $\nu$.
We denote the two possible values it can take by $\gamma$ and $\delta$ where $\gamma\neq \delta$.
Then we can rewrite the above,
\es{}{
F_{\mu\rho}\tilde{F}^{\nu\rho} = {1\over 2}F_{\mu\gamma}\epsilon^{\nu\gamma\alpha\beta} F_{\alpha\beta} + {1\over 2}F_{\mu\delta}\epsilon^{\nu\delta\alpha\beta} F_{\alpha\beta}~~[\gamma~{\rm and}~\delta~{\text{are not summed over}}].
}
The indices $\alpha$ and $\beta$ can then take values between $\mu$, $\delta$, and $\gamma$:
\es{}{
F_{\mu\rho}\tilde{F}^{\nu\rho} &= F_{\mu\gamma}\epsilon^{\nu\gamma\mu\delta} F_{\mu\delta} + F_{\mu\delta}\epsilon^{\nu\delta\mu\gamma} F_{\mu\gamma},~~[{\text{no summation over any index}}]\\
& = F_{\mu\gamma}F_{\mu\delta} (\epsilon^{\nu\gamma\mu\delta} + \epsilon^{\nu\delta\mu\gamma}) = 0.
}
Therefore, ${\cal O}_{8,7}$ does not contribute. 

The operator ${\cal O}_{8,8}$ is reducible.
To see this we can write,
\es{}{
\tilde{F}_{\mu\rho} \tilde{F}^{\nu\rho} &= {1\over 4} \epsilon_{\mu\rho\alpha\beta} \epsilon^{\nu\rho\gamma\delta} F^{\alpha\beta} F_{\gamma\delta}  = {1\over 2} \left(F^{\gamma\delta} F_{\gamma\delta} \delta^{\nu}_\mu - 2 F^{\nu\delta} F_{\mu\delta} \right).
}
This implies ${\cal O}_{8,8}$ is reducible to ${\cal O}_{8,1}$ and ${\cal O}_{8,6}$.

Next we focus on ${\cal O}_{8,9}$ which can be written as
\es{}{
{\cal O}_{8,9} = \nabla^{\mu}({\cal H}^\dagger {\cal H}) \nabla_\mu\nabla_\nu\phi \nabla^\nu\phi = -{1\over 2}\square({\cal H}^\dagger {\cal H})\nabla_\nu\phi \nabla^\nu\phi +{\rm boundary~term}.
}
where we have used IBP in the last step. Using methods similar to those used earlier, we can show the boundary term scales as $\eta_0^{1/2}\rightarrow 0$, and hence is not relevant for cosmological correlators.
The surviving term, however, is equivalent to ${\cal O}_{8,4}$.
To see this, we can use the EOM~\eqref{eq:eom_H} for ${\cal H}$ and $\nabla_\mu A^\mu =0$ to write,
\es{}{
\square({\cal H}^\dagger {\cal H}) = 2 (D_\mu {\cal H})^\dagger D^\mu {\cal H} + 2 V'(|{\cal H}|^2) |{\cal H}|^2.
}
The last term determined by the Higgs potential contributes to ${\cal O}_{6,1}$ and ${\cal O}_{8,3}$. 

Finally, ${\cal O}_{8,10}$ can be reduced using similar techniques.
First, we can rewrite it using the EOM~\eqref{eq:eom_A} and an IBP,
\es{}{
{1\over 2g_A}\nabla_\rho F^{\rho\mu} \nabla_\mu\left( \nabla_\nu \phi \nabla^\nu \phi\right) = -{1\over 2g_A}\nabla_\mu\nabla_\rho F^{\rho\mu} \left( \nabla_\nu \phi \nabla^\nu \phi\right) + {\rm boundary~term}.
}
Similar as above, one can check that the boundary term vanishes as $\eta_0^{3/2}$ at late times.
Noting that $\nabla_\mu\nabla_\rho F^{\rho\mu} \propto R_{\mu\rho}F^{\rho\mu} = 0$, we conclude that ${\cal O}_{8,10}$ does not contribute. Thus we find that there are six non-redundant operators at dimension 8.

\subsection{Dimension 9}
At dimension 9 we start with a set of operators summarized in Table~\ref{tab:dim9}.
\begin{table}
	\begin{center}
  \begin{tabular}{ | c | c |}
    \hline
    Operator & Expression \\ \hline
${\cal O}_{9,1}$ & $\nabla_\mu\phi ({\cal H}^\dagger D^\mu {\cal H} +  (D^\mu {\cal H})^\dagger {\cal H})|{\cal H}|^4 = \nabla_\mu\phi \nabla^\mu({\cal H}^\dagger {\cal H}) |{\cal H}|^4$\\
\hline
${\cal O}_{9,2}$ & $\nabla_\mu\phi (-i)({\cal H}^\dagger D^\mu {\cal H} -  (D^\mu {\cal H})^\dagger {\cal H}) |{\cal H}|^4 = {1\over g}\nabla_\mu\phi \nabla_\nu F^{\nu\mu} |{\cal H}|^4$\\
\hline
${\cal O}_{9,3}$ & $\nabla^\nu\phi ({\cal H}^\dagger D^\mu {\cal H} +  (D^\mu {\cal H})^\dagger {\cal H}) |{\cal H}|^2 F_{\mu\nu} = \nabla^\nu\phi \nabla^\mu({\cal H}^\dagger {\cal H}) |{\cal H}|^2 F_{\mu\nu}$\\
\hline
${\cal O}_{9,4}$ & $\nabla^\nu\phi (-i)({\cal H}^\dagger D^\mu {\cal H} -  (D^\mu {\cal H})^\dagger {\cal H}) |{\cal H}|^2 F_{\mu\nu} = {1\over g} \nabla^\nu\phi \nabla_\alpha F^{\alpha\mu} |{\cal H}|^2 F_{\mu\nu}$\\
\hline
${\cal O}_{9,5}$ & $\nabla_\nu\phi ({\cal H}^\dagger D^\mu {\cal H} +  (D^\mu {\cal H})^\dagger {\cal H}) F_{\mu\alpha} F^{\nu \alpha} = \nabla_\nu\phi \nabla^\mu({\cal H}^\dagger {\cal H}) F_{\mu\alpha} F^{\nu \alpha}$\\
\hline
${\cal O}_{9,6}$ & $\nabla_\mu\phi ({\cal H}^\dagger D^\mu {\cal H} +  (D^\mu {\cal H})^\dagger {\cal H}) F_{\alpha\nu} F^{\alpha\nu} = \nabla_\mu\phi \nabla^\mu({\cal H}^\dagger {\cal H}) F_{\alpha\nu} F^{\alpha\nu}$\\
\hline
${\cal O}_{9,7}$ & $\nabla_\mu\phi ({\cal H}^\dagger D^\mu {\cal H} +  (D^\mu {\cal H})^\dagger {\cal H}) F_{\alpha\nu} \tilde{F}^{\alpha\nu} = \nabla_\mu\phi \nabla^\mu({\cal H}^\dagger {\cal H}) F_{\alpha\nu} \tilde{F}^{\alpha\nu}$\\
\hline
${\cal O}_{9,8}$ & $\nabla_\nu\phi (-i)({\cal H}^\dagger D^\mu {\cal H} -  (D^\mu {\cal H})^\dagger {\cal H}) F_{\mu\alpha} F^{\nu \alpha} = {1\over g}\nabla_\nu\phi \nabla_\beta F^{\beta \mu} F_{\mu\alpha} F^{\nu \alpha}$\\
\hline
${\cal O}_{9,9}$ & $\nabla_\mu\phi (-i)({\cal H}^\dagger D^\mu {\cal H} -  (D^\mu {\cal H})^\dagger {\cal H}) F_{\alpha\nu} F^{\alpha\nu} = {1\over g} \nabla_\mu\phi \nabla_\beta F^{\beta\mu} F_{\alpha\nu} F^{\alpha\nu}$\\
\hline
${\cal O}_{9,10}$ & $\nabla_\mu\phi (-i)({\cal H}^\dagger D^\mu {\cal H} -  (D^\mu {\cal H})^\dagger {\cal H}) F_{\alpha\nu} \tilde{F}^{\alpha\nu} = {1\over g} \nabla_\mu\phi \nabla_\beta F^{\beta\mu} F_{\alpha\nu} \tilde{F}^{\alpha\nu}$\\
\hline
${\cal O}_{9,11}$ & $\nabla_\mu\phi ({\cal H}^\dagger D^\mu {\cal H} +  (D^\mu {\cal H})^\dagger {\cal H}) (\nabla_\nu\phi)^2 = \nabla_\mu\phi \nabla^\mu({\cal H}^\dagger {\cal H}) (\nabla_\nu\phi)^2$\\
\hline
${\cal O}_{9,12}$ & $\nabla_\mu\phi (-i)({\cal H}^\dagger D^\mu {\cal H} -  (D^\mu {\cal H})^\dagger {\cal H}) (\nabla_\nu\phi)^2 =  {1\over g} \nabla_\mu\phi\nabla_\alpha F^{\alpha\mu} (\nabla_\nu\phi)^2$\\
\hline
${\cal O}_{9,13}$ & $\nabla^\nu\phi \nabla_\nu({\cal H}^\dagger {\cal H}) |D_\mu {\cal H}|^2$\\
\hline
${\cal O}_{9,14}$ & $\nabla_\mu\phi \nabla^\nu({\cal H}^\dagger {\cal H}) (D^\mu {\cal H})^\dagger D_\nu {\cal H}$\\
\hline
${\cal O}_{9,15}$ & $\nabla_\nu\phi \nabla_\alpha F^{\alpha\nu} |D_\mu {\cal H}|^2$\\
\hline
${\cal O}_{9,16}$ & $\nabla_\nu\phi \nabla_\alpha F^{\alpha\mu}(D^\nu {\cal H})^\dagger D_\mu {\cal H}$\\
\hline
${\cal O}_{9,17}$ & $\nabla_\nu\nabla_\mu\phi \nabla^\mu({\cal H}^\dagger {\cal H}) \nabla^\nu({\cal H}^\dagger {\cal H})$\\
\hline
${\cal O}_{9,18}$ & $\nabla_\nu\nabla_\mu\phi \nabla_\alpha F^{\alpha\mu}\nabla_\beta F^{\beta\nu}$\\
\hline
${\cal O}_{9,19}$ & $\nabla_\nu\nabla_\mu\phi \nabla_\alpha F^{\alpha\mu}\nabla^\nu({\cal H}^\dagger {\cal H})$\\
\hline
${\cal O}_{9,20}$ & $\nabla_\nu\nabla_\mu\phi \nabla^\mu F^{\rho\nu}\nabla_\rho({\cal H}^\dagger {\cal H})$\\
\hline
\end{tabular}
\end{center}
\caption{Allowed operators at dimension 9.}
\label{tab:dim9}
\end{table}
Among these, ${\cal O}_{9,1}$ does not contribute, as can be seen by doing an IBP which gives a vanishing boundary term, along with a term involving $\square\phi$ which vanishes in the slow-roll limit.
Using IBP, we can also check that ${\cal O}_{9,2}$ and ${\cal O}_{9,3}$ are equivalent.
${\cal O}_{9,17}$ and ${\cal O}_{9,20}$ are both reducible in terms of other operators.
There are a priori other permutations with three derivatives acting on $\phi$.
Those terms can, however, be reduced by using the fact that for a maximally symmetric spacetime such as dS, we can write the Riemann tensor as $R_{\mu\nu\rho\sigma} \propto (g_{\mu\rho} g_{\nu\sigma} - g_{\mu\sigma} g_{\nu\rho})$.
The Bianchi identity for $F_{\mu\nu}$ is also useful in reducing certain terms.
All the other terms would contribute to cosmological correlators, albeit with suppressed contributions compared to operators at lower dimensions, as we will see in the next section.

However, the operator ${\cal O}_{9,12}$ is special since it gives rise to a quadratic mixing between the inflaton and the longitudinal gauge boson.
Such an operator is observationally relevant since it would mediate tree-level NG, and we have seen that no other operator up to dimension 8 could give rise to such a mixing.
To elaborate on this further, we can rewrite ${\cal O}_{9,12}$ after an IBP as, (dropping the gauge coupling)
\es{eq:O91_simplified}{
{\cal O}_{9,12} =\nabla_\nu\left[(\nabla_\rho \phi \nabla^\rho \phi) \nabla_\mu \phi F^{\nu\mu}\right] - \nabla_\nu (\nabla_\rho \phi \nabla^\rho \phi) \nabla_\mu \phi F^{\nu\mu}.
}
Here we have used the fact that $F^{\nu\mu}\nabla_\mu\nabla_\nu \phi=0$.
\paragraph{The Boundary Term.}
We first consider the boundary term, following an analysis similar to the above.
We can write the boundary term as,
\es{}{
{i\over g_A}\int \D^3x \sqrt{|\gamma|}n_\nu \left[(\nabla_\rho \phi \nabla^\rho \phi) \nabla_\mu \phi F^{\nu\mu}\right]
}
Therefore, at late times this term scales as,
\es{}{
\sim {1\over \eta_0^3}\times {1\over \eta_0} \times \eta_0^2 \partial_i\phi \times \eta_0^4 F_{\eta i} \rightarrow 0.
}
Thus this term can be dropped.
\paragraph{The Bulk Term.}
To obtain the inflaton-gauge boson mixing, we can focus on the $F^{\eta i}$ component from~\eqref{eq:O91_simplified}.
Since our main purpose to illustrate the form of the quadratic mixing, we will not track the overall numerical and $\eta$ factors.
If we set $\mu=i$ and $\nu=0$, then the term would have a $\ddot{\phi}_0$, and hence it would be slow-roll suppressed.
However, for $\mu=0$ and $\nu=i$, we would have a contribution which is quadratic in fluctuations,
\es{}{
{\cal O}_{9,12} \supset \dot{\phi}_0^2\partial_i\dot{\phi} F^{i\eta}.
}
To simplify this further we can do a spatial IBP to write,
\es{}{
{\cal O}_{9,12} \supset \dot{\phi}_0^2\dot{\phi} \partial_iF^{i\eta}.
}
and use~\eqref{eq:AconEq} and~\eqref{eq:AetaEOM} to write
\es{}{
{\cal O}_{9,12} \supset \dot{\phi}_0^2m^2 \dot{\xi}  A_\eta.
}
Here we have written the inhomogeneous part of the inflaton field: $\phi(t,\vec{x}) = \phi_0(t) + \xi(t,\vec{x})$.
This matches with the conclusion in~\cite{Kumar:2017ecc}, however only this particular dimension 9 operator was considered in isolation.

\subsection{Summary and Classification}

The non-redundant operators up to dimension-9 are summarized in Table~\ref{table:summary}. In the third column, we indicate whether the operator contributes to the bispectrum at tree or loop level. This differs depending on the phase of the gauge theory. For example, $\mathcal{O}_{6,1}$ can clearly contribute to the bispectrum at tree level in the broken phase, when we can decompose the Higgs field in unitary gauge as ${\cal H} =  \left(h+v \right) /\sqrt{2}$, with $h$ an interacting degree of freedom and $v$ the VEV. In the unbroken phase, this is not possible, and the operator may only contribute via a loop diagram. 

On top of the tree/loop classification of each operator's contribution to observables, in weakly coupled UV completions it may also be possible to assign a tree/loop classification to the operator's Wilson coefficients~\cite{Einhorn:2013kja}. At present the tree/loop classification of Wilson coefficients has only been extended through dimension-8~\cite{Craig:2019wmo}, and we do not explicitly classify Wilson coefficients here.

\begin{table}
	\begin{center}
  \begin{tabular}{ | c | c | c |}
    \hline
    {\bf Dimension} & {\bf Operator}  & {\bf Observables} \\ \hline
    \hline
    5 & \textcolor{red}{${\cal O}_{5,4}=\phi F_{\mu\nu} \tilde{F}^{\mu\nu}$}
     & Loop~\cite{Wang:2020ioa}  \\ 
   \hline \hline
    6 & \textcolor{red}{${\cal O}_{6,1}=(\nabla_\mu\phi)^2 {\cal H}^\dagger {\cal H}$} 
     & \textcolor{blue}{Tree}~\cite{Kumar:2017ecc} and Loop~\cite{Chen:2016hrz} \\ \hline \hline 
    7 & \textcolor{red}{${\cal O}_{7,2} =|{\cal H}|^2 \nabla_\mu \phi \nabla_\nu F^{\nu\mu}$} & Loop \\ 
     & ${\cal O}_{7,4} = F_{\mu\nu}\nabla^\mu\phi \nabla_\rho F^{\rho \nu}$ & Loop  \\ \hline
    \hline
    8 & \textcolor{red}{$\mathcal{O}_{8,1} = F_{\mu \nu} F^{\mu \nu} (\nabla_\rho \phi)^2$} & Loop~\cite{Chen:2016hrz}  \\
      & \textcolor{red}{$\mathcal{O}_{8,2} = F_{\mu \nu} \tilde F^{\mu \nu} (\nabla_\rho \phi)^2$} & Loop  \\
      & $\mathcal{O}_{8,3} = |{\cal H}|^4 (\nabla_\mu \phi)^2$ & \textcolor{blue}{Tree} and Loop  \\
      & $\mathcal{O}_{8,4} = |D_\mu {\cal H}|^2 (\nabla_\nu \phi)^2$  & Loop~\cite{Chen:2016hrz} \\
      & $\mathcal{O}_{8,5} = (D^\mu {\cal H})^\dag D^\nu {\cal H} \nabla_\mu \phi \nabla_\nu \phi$  & Loop \\
 & \textcolor{red}{$\mathcal{O}_{8,6} = F_{\mu \rho} F^{\nu \rho} \nabla^\mu \phi \nabla_\nu \phi$} & Loop \\
    \hline
    \hline
    9 & ${\cal O}_{9,2} = |{\cal H}|^2 {\cal O}_{7,2}$  & Loop  \\
    & ${\cal O}_{9,4} = |{\cal H}|^2 {\cal O}_{7,4}$ & Loop \\
    & ${\cal O}_{9,5} = \nabla_\nu\phi \nabla^\mu({\cal H}^\dagger {\cal H}) F_{\mu\alpha} F^{\nu \alpha}$ & Loop  \\
    & ${\cal O}_{9,6} = {\cal O}_{5,1} F_{\alpha\nu} F^{\alpha\nu}$ & Loop  \\
    & ${\cal O}_{9,7} = {\cal O}_{5,1} F_{\alpha\nu} \tilde{F}^{\alpha\nu}$ & Loop \\
    & ${\cal O}_{9,8} = \nabla_\nu\phi \nabla_\beta F^{\beta \mu} F_{\mu\alpha} F^{\nu \alpha}$ & Loop  \\
    & ${\cal O}_{9,9} = {\cal O}_{5,3} F_{\alpha\nu} F^{\alpha\nu}$ & Loop\\
    & ${\cal O}_{9,10} = {\cal O}_{5,3} F_{\alpha\nu} \tilde{F}^{\alpha\nu}$ & Loop \\
    & ${\cal O}_{9,11} = {\cal O}_{5,1} (\nabla_\mu\phi)^2$ & \textcolor{blue}{Tree} and Loop   \\
    & \textcolor{red}{${\cal O}_{9,12} = {\cal O}_{5,3} (\nabla_\mu\phi)^2$} & {\color{blue} Tree} \cite{Kumar:2017ecc} and Loop   \\
    & ${\cal O}_{9,13} = {\cal O}_{5,1} |D_\mu {\cal H}|^2$ & Loop  \\
    & ${\cal O}_{9,14} = \nabla_\mu\phi \nabla^\nu({\cal H}^\dagger {\cal H}) (D^\mu {\cal H})^\dagger D_\nu {\cal H}$  & Loop  \\
    & ${\cal O}_{9,15} = {\cal O}_{5,3} |D_\mu {\cal H}|^2$  & Loop  \\
    & ${\cal O}_{9,16} = \nabla_\nu\phi \nabla_\alpha F^{\alpha\mu}(D^\nu {\cal H})^\dagger D_\mu {\cal H}$  & Loop \\
& ${\cal O}_{9,18} = \nabla_\nu\nabla_\mu\phi \nabla_\alpha F^{\alpha\mu}\nabla_\beta F^{\beta\nu}$ & Loop \\
& ${\cal O}_{9,19} = \nabla_\nu\nabla_\mu\phi \nabla_\alpha F^{\alpha\mu}\nabla^\nu({\cal H}^\dagger {\cal H})$ & Loop \\
    \hline
  \end{tabular}
\end{center}
\caption{A minimal operator basis up to dimension-9. We have dropped some overall prefactors to write the operators more compactly in terms of dimension-5 and -7 operators. Operators with leading effects (as described in Sec.~\ref{sec:obs}) are highlighted in red. The third column indicates whether these operators contribute to the bispectrum at tree or loop level. Observables highlighted in blue only arise in the broken phase.}
\label{table:summary}
\end{table}

\section{Observational Implications}\label{sec:obs}
Having constructed a minimal basis, we now study the predictions for NG, especially in the context of the cosmological collider.
We first briefly review some aspects that will also set up the notation.
As discussed in the Introduction, particles with masses of order $H_{\rm inf}$ can be produced as the Universe inflates.
After production these particles can propagate on-shell, oscillating in time, and eventually decay into inflaton fluctuations.
Such processes then give rise to non-trivial correlations among different inflaton fluctuations, in particular, three- and higher-point correlation functions.
In this work, we will focus on the three-point function, i.e., the bispectrum, characterized by three spatial momenta $\vec{k}_1, \vec{k}_2, \vec{k}_3$:
\es{}{
\langle {\cal R}(\vec{k}_1) {\cal R}(\vec{k}_2) {\cal R}(\vec{k}_3)  \rangle = (2\pi)^3 \delta(\vec{k}_1+\vec{k}_2+\vec{k}_3) B(k_1, k_2, k_3).
}
The $\delta$ function above enforces spatial momentum conservation.
We have denoted the gauge invariant comoving curvature perturbation by ${\cal R}$; for a detailed definition and review see, e.g.,~\cite{Malik:2008im}.  
Conventionally, the function $B$ is normalized with respect to the power spectrum so that there is no overall scale dependence,
\es{}{
F(k_1, k_2, k_3) = {5 \over 6} {B(k_1, k_2, k_3) \over P_{\cal R}(k_1) P_{\cal R}(k_2) + P_{\cal R}(k_2) P_{\cal R}(k_3) + P_{\cal R}(k_1) P_{\cal R}(k_3)}.
}
It is also a convention to characterize the `strength' of NG, at the equilateral configuration where $k_1=k_2=k_3$, in terms of a single number $f_{\rm NL} \equiv F(k,k,k)$.
In the case of the cosmological collider, the function $F$ exhibits oscillations as a function of $k_3/k_1$, especially in the squeezed limit $k_3 \ll k_1\approx k_2$.
We will denote the associated strength of NG by a parameter $f_{\rm osc}$, defined via:
\es{}{
F_{\rm sq} \approx {5 \over 12} {B(k_1, k_2, k_3) \over P_{\cal R}(k_1) P_{\cal R}(k_3)} \equiv |f_{\rm osc}|\left[ \left( {k_3 \over k_1}\right)^{{3\over 2} + i\mu} + {\rm c.c.}\right].
}
In the following, we will estimate the parametric dependence of $f_{\rm osc}$ on various Wilson coefficients and identify which operators are expected to give a leading signal in a generic EFT.
To that end, we briefly revisit the power counting scheme.

\subsection{Power Counting}
As discussed in Section \ref{sec:basis}, the power counting scheme for the gauge-Higgs-inflaton EFT can be organized in terms of operator dimension. For simplicity we take operators to be suppressed by appropriate powers of a common UV scale $\Lambda$ with Wilson coefficients $c_{n,a}$,
\es{}{
{\cal L} \supset \sum_{n=5,\cdots} \frac{c_{n,a}}{\Lambda^{n-4}} \mathcal{O}_{n,a}.
}
Here $n$ determines the dimension of the operator while the index $a$ runs over all the operators having the same dimension.
The EFT scale has to satisfy certain restrictions.
To control the inflaton derivative expansion in $(\partial \phi)^2/\Lambda^4$, we require $\Lambda >\sqrt{\dot{\phi}}_0$~\cite{Creminelli:2003iq}.
We also require $\Lambda > v$ to control the expansion in $v^2/\Lambda^2$, i.e., for the EFT to be suitably organized in terms of the linearly-realized gauge symmetry.

\subsection{`Monochromatic' Operators}
The operators summarized in Table~\ref{table:summary} contribute to several types of cosmological correlators.
A given diagram could involve either the Higgs or the gauge boson or both.
Diagrams in which both the Higgs and the gauge boson are present can give rise to interesting signatures.
For a recent study outlining the techniques for computing such diagrams involving more than one massive field, see~\cite{Xianyu:2023ytd}.
However, the `monochromatic' signatures that could let us extract the mass and spin of the boson in the most immediate manner would involve either the Higgs or the gauge boson, but not both.\footnote{If there is a hierarchy between the Higgs and the gauge boson mass, we can integrate out the heavier particle to effectively obtain a monochromatic contribution for the other particle.
Here we instead focus on the case where both the Higgs mass and the gauge boson mass are comparable.}
With this in mind, we now highlight which operators would give rise to such monochromatic signatures. 

Note that a given monochromatic signature typically accumulates contributions from multiple operators at successive orders in power counting, where higher-order contributions are suppressed by appropriate powers of $v/\Lambda$ or $H_{\rm inf}/\Lambda.$ In what follows, the {\it leading} monochromatic signatures are those that arise at the lowest order in power counting. Leading operators are indicated in red in Table~\ref{table:summary}.  

\subsubsection{Higgs Signature}
The leading monochromatic operator involving the Higgs arises at dimension 6, namely ${\cal O}_{6,1}$.
In the broken phase, this operator gives rise to tree-level NG, as was studied in detail in~\cite{Kumar:2017ecc}.
On the other hand, in the unbroken phase, the operator contributes to NG at the loop level, as studied in~\cite{Chen:2016hrz}.

At dimension 7 there are no monochromatic Higgs operators, since all the non-redundant operators involve gauge bosons as well.
At dimension 8, monochromatic Higgs signatures arise from ${\cal O}_{8,3}, {\cal O}_{8,4}, {\cal O}_{8,5}$.
However, we typically expect these contributions to be suppressed by additional powers of $(H_{\rm inf}/\Lambda)$ or $(v/\Lambda)$ compared to the dimension-6 contribution.
At dimension 9, ${\cal O}_{9,11}$, ${\cal O}_{9,13}$, ${\cal O}_{9,14}$ would contribute, with additional suppression by powers of $(H_{\rm inf}/\Lambda)$ and/or $(v/\Lambda)$.
In particular, ${\cal O}_{9,13}$ and ${\cal O}_{9,14}$ would contribute to vertices having at least three Higgs fluctuations.

\subsubsection{Gauge Boson Signature}

At loop-level, the leading monochromatic gauge boson signature arises from ${\cal O}_{5,4}$.
This operator has been studied extensively in the context of axion inflation~\cite{Anber:2009ua}, and in the context of the cosmological collider~\cite{Wang:2020ioa}.
The operator ${\cal O}_{7,4}$ can also contribute, but potentially without the chemical potential-like structure which arises from ${\cal O}_{5,4}$.
There are multiple possible contributions at dimension 8.
For example, the effects of ${\cal O}_{8,1}$ were computed in~\cite{Chen:2016hrz}, while ${\cal O}_{8,2}$ and ${\cal O}_{8,6}$ would also contribute at the same order in EFT power counting.

At dimension 9, ${\cal O}_{9,12}$ is special since it can give rise to a quadratic mixing between the inflaton and the longitudinal gauge boson, as discussed above.
The operator ${\cal O}_{9,18}$ also contributes, albeit suppressed by powers of $(H_{\rm inf}/\Lambda)$. There are other operators at dimension 9 that involve one inflaton with three gauge bosons, and therefore do not contribute to a three-point function at the one-loop level.

\subsection{Estimates}
\begin{figure}
	\begin{center}
		\includegraphics[width=0.9\textwidth]{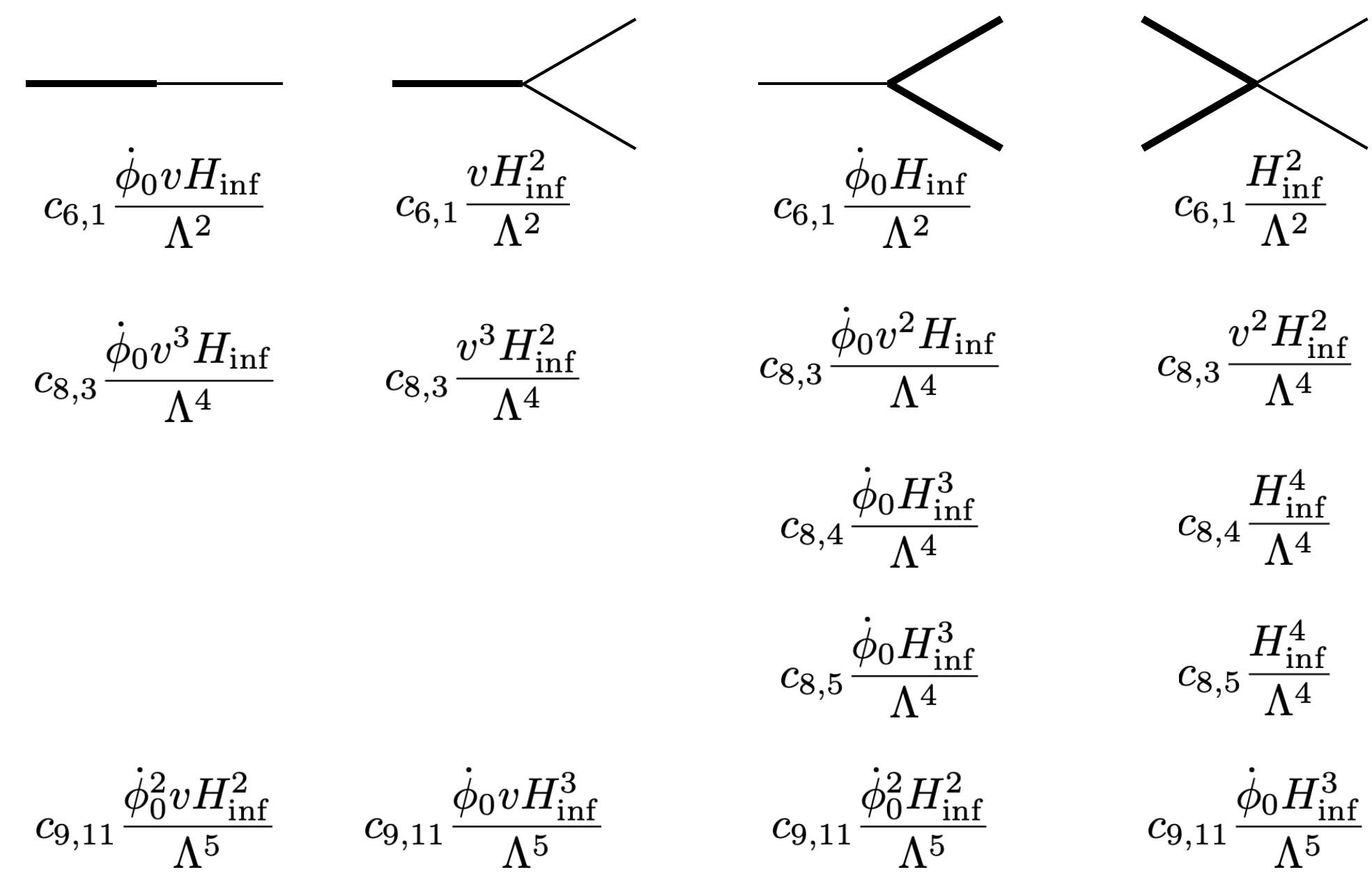}
	\end{center}
	\caption{Higgs-inflaton vertices from various operators present in Table~\ref{table:summary}. The inflaton (Higgs) is denoted by a thin (solid) line. We have not included vertices with more than two Higgs fluctuations.}
 \label{fig_Higgs}
\end{figure}

\begin{figure}
	\begin{center}
		\includegraphics[width=0.9\textwidth]{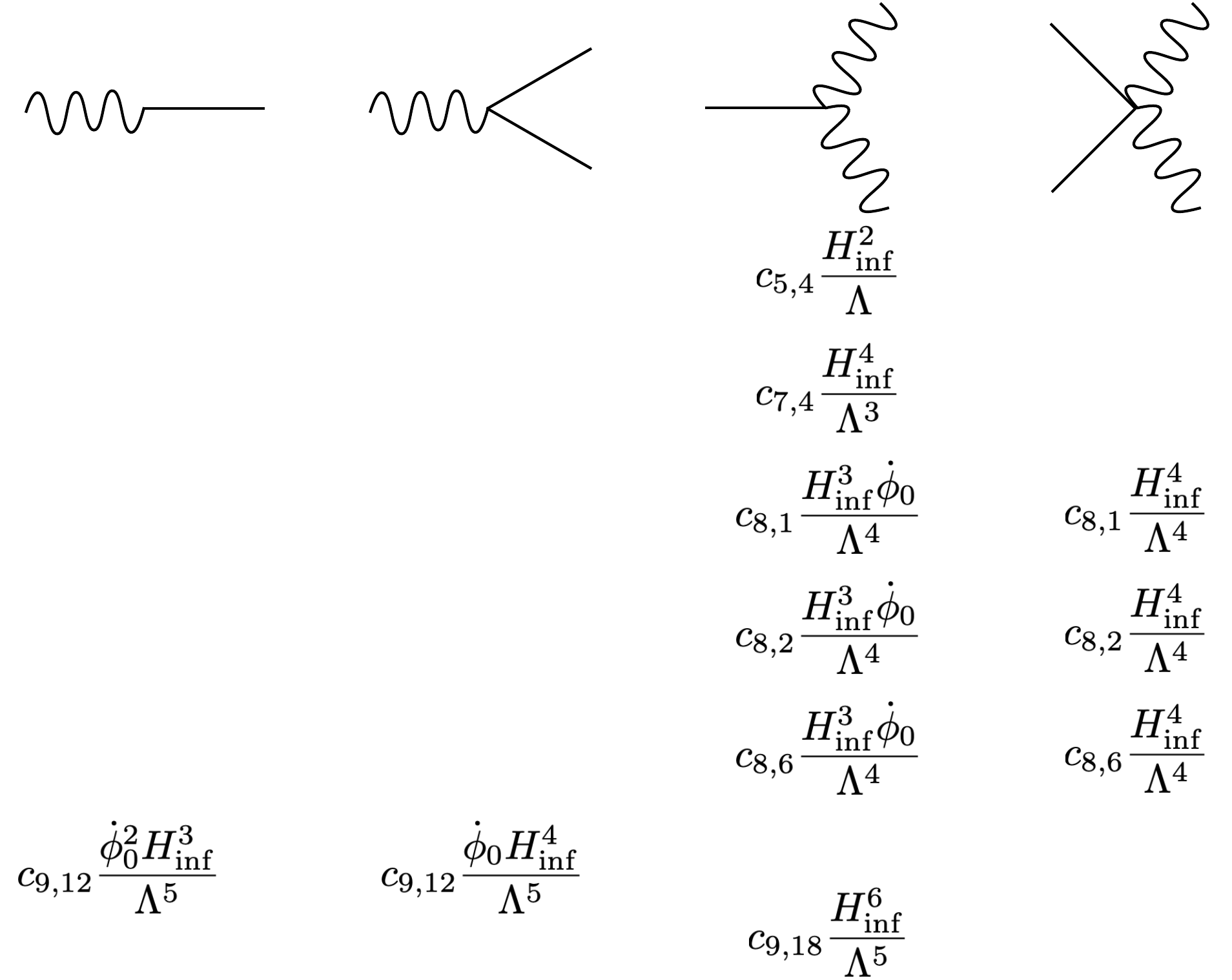}
	\end{center}
	\caption{Gauge boson-inflaton vertices from various operators present in Table~\ref{table:summary}. The inflaton (gauge boson) is denoted by a straight (wavy) line. We have not included vertices with more than two gauge bosons.}
 \label{fig_gauge}
\end{figure}

The vertices relevant for NG mediated at tree level or at one loop are summarized in Figs.~\ref{fig_Higgs} and~\ref{fig_gauge}.
There are four types of vertices, many of which accumulate contributions from more than one operator in Table~\ref{table:summary}, as indicated by the corresponding Wilson coefficients $c_{n,a}$.
For operators involving derivatives, we have estimated the size of the derivatives to be of order $H_{\rm inf}$, as appropriate for scenarios where all the mass scales are of the order $H_{\rm inf}$.

Using these vertices we can construct various tree and loop-level diagrams that can mediate non-gaussianities.
For illustration, we only consider the leading Higgs and gauge-boson mediated NG.
\paragraph{Higgs.}
The signature in the broken phase was discussed in~\cite{Kumar:2017ecc}; here we summarize the main conclusions.
The dimension 6 operator ${\cal O}_{6,1}$ would mediate the leading tree-level NG through the vertices shown on the first row of Fig.~\ref{fig_Higgs}.
The parametric dependence of NG from the so-called `single exchange' diagram can be estimated as,
\es{}{
f_{\rm osc, tree}^{\rm Higgs} \sim {\mu^4 v^2 \over \dot{\phi}_0^2 H_{\rm inf}^2}, 
}
where $\mu^2 \sim c_{6,1} \dot{\phi}_0^2/\Lambda^2$ denotes a `classical' correction to the Higgs mass from ${\cal O}_{6,1}$ when the inflaton is set to its VEV~\cite{Kumar:2017ecc}.
The total Higgs mass is then given by $m_{h, \rm full}^2 \sim \mu^2 + m_{h, \rm bare}^2$ where $m_{h, \rm bare}^2$ is the Higgs mass in the absence of any inflaton correction.
Since cosmological collider signatures would be exponentially suppressed for $m_{h, \rm full} \gg H_{\rm inf}$, we require $m_{h, \rm full}^2 \sim H_{\rm inf}^2$.
Barring any fine-tuning, this would mean $\mu^2 \sim H_{\rm inf}^2$ and $m_{h, \rm bare}^2 \sim H_{\rm inf}^2$.
This implies the going rate for NG is $f_{\rm osc, tree}^{\rm Higgs} \sim (H_{\rm inf}^2 v^2)/\dot{\phi}_0^2$.
For a more detailed numerical computation of the non-gaussianity, see Ref.~\cite{Kumar:2017ecc} where the exponential fall off of the NG as a function of increasing Higgs mass is also computed.

In the unbroken phase, the NG signature arises at one loop, mediated by the right two vertices in the first row of Fig.~\ref{fig_Higgs}.
The corresponding estimate is
\es{}{
f_{\rm osc, loop}^{\rm Higgs} \sim {1 \over 16\pi^2} {\mu^4 \over \dot{\phi}_0^2}.
}
For a detailed evaluation see Ref.~\cite{Chen:2016hrz}.

\paragraph{Gauge Boson.}
We first discuss the tree-level signature in the broken phase, which was also discussed in~\cite{Kumar:2017ecc}.
For a tree-level signature, an essential ingredient is a quadratic mixing between an inflaton and the longitudinal mode of the gauge boson.
Such a mixing arises at dimension 9, namely via ${\cal O}_{9,12}$, which also gives a cubic interaction between the gauge boson and the inflaton.
These two vertices can contribute to NG via the so-called single exchange diagram:
\es{}{
f_{\rm osc, tree}^{\rm gauge} \sim c_{9,12}^2 {\dot{\phi}_0^4 H^2 \over \Lambda^{10}}.
}
For $\Lambda \gtrsim \sqrt{\dot{\phi}_0}$, the above becomes $f_{\rm osc, tree}^{\rm gauge} \lesssim c_{9,12}^2 (H^2 / \dot{\phi}_0)$.
For a detailed evaluation see Ref.~\cite{Kumar:2017ecc}.
There is another diagram that can contribute to NG, involving the vertices with $c_{7,4}$ and $c_{9,12}$.
The strength can be estimated as,
\es{}{
f_{\rm osc,tree}^{\rm gauge} \sim c_{7,4} c_{9,12}^2 {\dot{\phi}_0^5 \over \Lambda^{13}}.
}
Taking $\Lambda \gtrsim \sqrt{\dot{\phi}_0}$, the above estimate becomes $f_{\rm osc,tree}^{\rm gauge} \lesssim c_{7,4} c_{9,12}^2 (H /\dot{\phi}_0^{3/2})$, and therefore this contribution is expected to be subdominant compared to the previous process mediated purely via $c_{9,12}$.
The dimension-5 operator determined by $c_{5,4}$ can give larger signals, both because it is a leading operator from an EFT perspective, and also because it can give a `chemical potential' for gauge boson, potentially leading to exponential particle production.
The cosmological collider signatures were computed in~\cite{Wang:2020ioa}.

\section{Conclusion}\label{sec:conc}

A systematic approach to constructing local EFTs entails not only fixing the power-counting and enumerating operators consistent with the infrared symmetries and fields, but also accommodating the resulting redundancy of description. This systematic approach is well-established in flat-space EFTs, where the irrelevance of boundary terms and invariance of $S$-matrix elements under field redefinitions make operator redundancies transparent. The situation is more complicated in cosmological EFTs, where boundary terms are not always negligible and the observables of interest are sensitive to field redefinitions. While minimal operator bases for inflaton self-interactions have been enumerated in various inflationary EFTs, much less progress has been made for EFTs of heavy particles coupled to the inflaton beyond the simplest cases at lowest order. 

In this paper, we have developed a minimal operator basis for an abelian gauge-Higgs-inflaton EFT up to dimension 9, an archetypal example of a sector of heavy fields coupled to the inflaton relevant for cosmological collider physics. We have identified low-dimensional operators that are entirely redundant, as well as new non-redundant operators with potentially interesting observational signatures. Along the way, we have identified a number of useful methods for checking boundary terms arising from IBP relations, which can readily be applied to other EFTs of heavy particles coupled to the inflaton. The systematic enumeration of minimal operator bases in these EFTs is invaluable in light of the considerable interest in their cosmological signatures.

There are, of course, a number of interesting future directions. The methods presented in this paper may readily be generalized to other sectors coupled to the inflaton, including fermions and non-abelian gauge bosons. While we have focused on a Lorentz-preserving EFT of inflation with a shift-symmetric inflaton, similar methods may be applied in the more general Goldstone EFT of inflation~\cite{Cheung:2007st}. Further, we have enumerated a number of operators with observational effects at loop-order. The precise computation of these effects remains important and will be the subject of future work. Finally, it would be very interesting to extend the general methods developed for operator bases in flat space~\cite{Henning:2017fpj} to cosmological contexts by accounting for the possible role of boundary terms. More broadly, we hope that the extensive attention devoted to operator bases in flat space EFTs may be equally applied to the plethora of EFTs arising in cosmological settings.

\section*{Acknowledgements}
We would like to thank LianTao Wang for useful conversations, Daniel Green for comments on the manuscript, and Dave Sutherland for his deft help with \texttt{DEFT}.
The work of NC and AM was supported in part by the Department of Energy under grant DE-SC0011702. The work of SK was supported in part by NSF grant PHY-2210498 and the Simons Foundation. AM acknowledges the hospitality of the Berkeley Center for Theoretical Physics and the theory group of the Lawrence Berkeley National Laboratory. The research of AM is supported by the U.S. Department
of Energy, Office of Science, Office of Workforce Development for Teachers and Scientists, Office of Science Graduate Student Research (SCGSR) program under contract
number DE-SC0014664.

\begin{appendix}
\section{IBP, EOM, and Field Redefinitions in dS: An Explicit Example}\label{app.detailed_example}
We are interested in computing cosmological correlation functions on a late time slice at $\eta_0$.
Therefore, the boundary terms that arise while employing IBP can potentially contribute to such correlation functions.
Here we investigate the nature of these boundary terms by focusing on a concrete example involving a massive field $\sigma$ having a mass $m$, defined in terms of the variable $\mu = (m^2/H^2 - 9/4)^{1/2}$.
In particular, we focus on the standard quasi-single field operator~\cite{Chen:2009zp}:
\es{}{
{\cal O}_1={1\over \Lambda}\int \D^4 x \sqrt{|g|}(\nabla_\mu\phi)(\nabla^\mu\phi)\sigma.
}
By IBP we can write this as,
\es{}{
{\cal O}_1= {1\over \Lambda}\int \D^4 x \sqrt{|g|}\left[\nabla_\mu (\phi \nabla^\mu\phi\cdot \sigma)- \phi\square \phi \cdot \sigma - \phi \nabla^\mu\phi \nabla_\mu\sigma\right].
}
The term involving $\square \phi$ does not contribute to in-in correlators by virtue of the inflaton EOM $\square\phi \approx 0$, where we ignore the contribution from the inflaton potential.
We rewrite the remaining terms as,
\es{eq:O2}{
{\cal O}_2&={1\over 2 \Lambda} \int \D^4 x \sqrt{|g|}\left[\nabla_\mu (\nabla^\mu(\phi^2)\cdot \sigma) - \nabla^\mu(\phi^2) \nabla_\mu\sigma\right] \\
&= {\cal O}_{2,\rm TD} + {\cal O}_{2,\rm Bulk}.
}
Here we have separated the total derivative (TD) and the bulk term. As noted in \cite{Assassi:2013gxa}, the boundary term can be neglected for equal-time correlation functions in the in-in formalism because such terms are associated with equal-time commutators and can be removed by a redefinition of the local operators. In what follows, we explore this in great detail. 

\begin{figure}
	\begin{center}
		\includegraphics[width=0.9\textwidth]{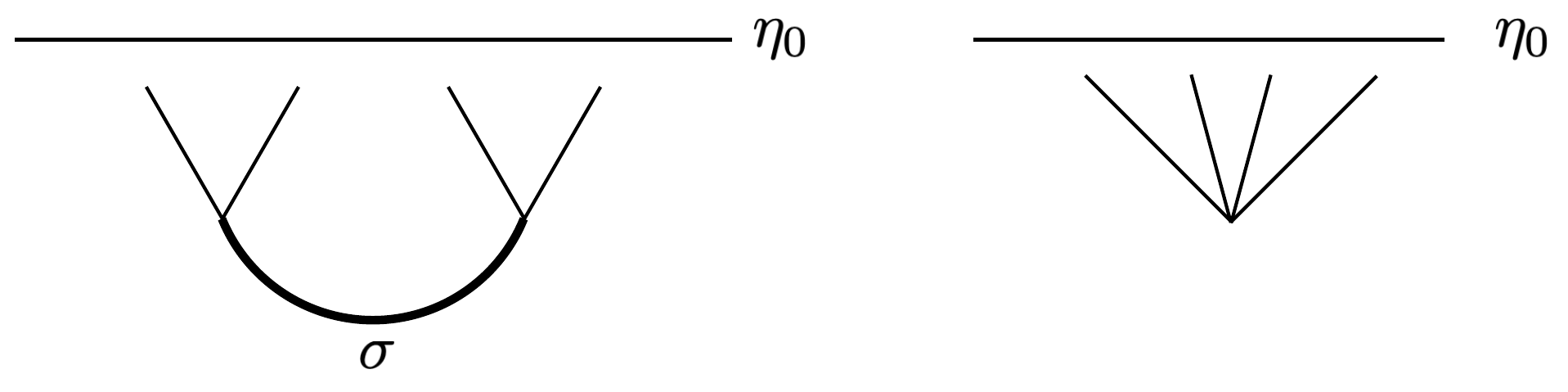}
	\end{center}
	\caption{{\it Left.} Four point function mediated by ${\cal O}_1$ and ${\cal O}_2$. {\it Right.} Contact interaction mediated by ${\cal O}_{2,\rm TD}$.}
 \label{fig_4pt}
\end{figure}

We first check that ${\cal O}_1$ and ${\cal O}_2$ give exactly the same contribution to in-in correlation functions, as they should.
Our goal would then be to understand the contribution from ${\cal O}_{2,\rm TD}$.
To that end, we focus on a four point function as shown in the left panel of Fig.~\ref{fig_4pt}.
There are four terms contributing to this four point function.
We call the associated contributions as, ${\cal I}_{++}$, ${\cal I}_{+-}$, ${\cal I}_{-+}$, and ${\cal I}_{--}$, of which the third and the fourth are the complex conjugates of the second and the first, respectively (the $+ (-)$ sign denotes that the vertex come from a time (anti-time) ordering operator).
Therefore, we only consider ${\cal I}_{++}$ and ${\cal I}_{+-}$ to show the equivalence of ${\cal O}_{1}$ and ${\cal O}_{2}$.
For convenience, we write the TD term as,
\es{eq:O_2TD}{
{\cal O}_{2,\rm TD} = {1\over 2\Lambda}\int \D^4 x \partial_\mu\left( \sqrt{|g|} \partial^\mu(\phi^2)\cdot \sigma\right) = -{1\over 2\Lambda}\int \D\eta \partial_\eta \left( {1\over \eta^2} \partial_\eta(\phi^2)\cdot \sigma\right).
}
In the last equality, we have dropped the spatial boundary terms, assuming fields decay at spatial infinity.
A similar operation can not be naively done for the temporal boundary, since we are interested in computing the correlation functions on the same boundary.
 
\begin{figure}
	\begin{center}
		\includegraphics[width=0.49\textwidth]{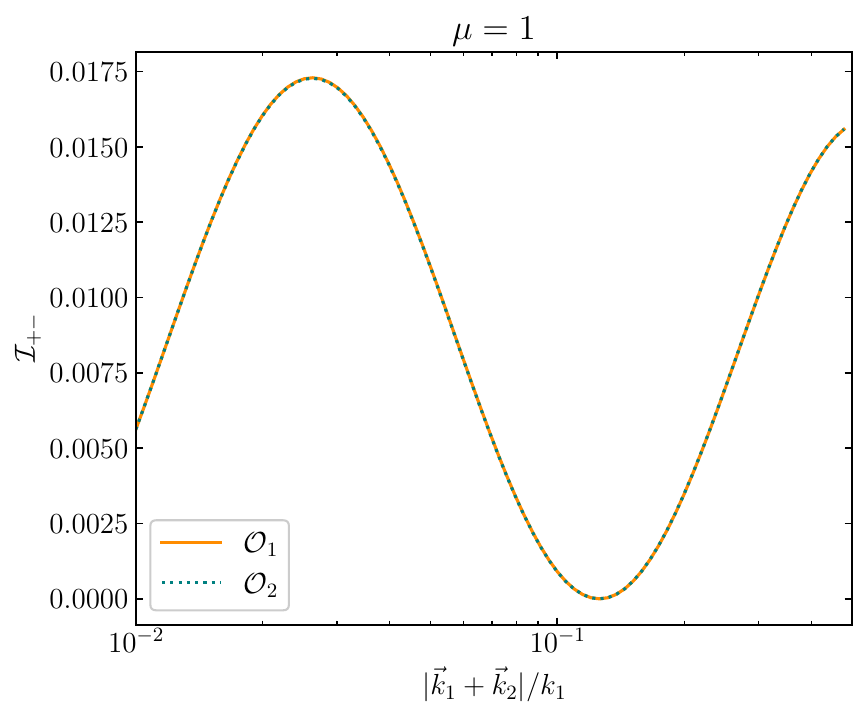}
		\includegraphics[width=0.49\textwidth]{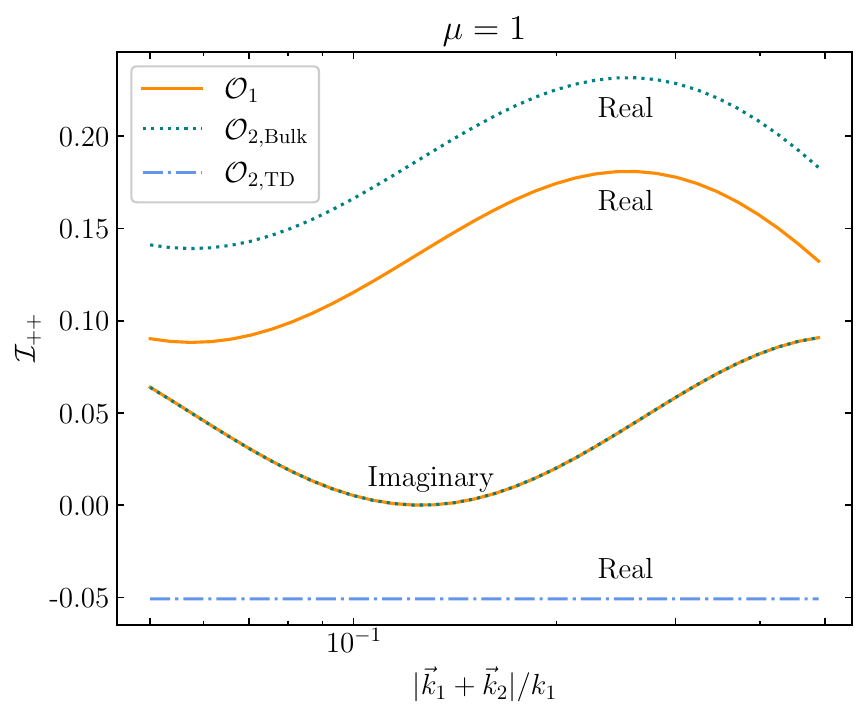}
	\end{center}
	\caption{Comparison between the contributions from ${\cal O}_1$ and ${\cal O}_2$.
	{\it Left.} The ${\cal I}_{+-}$ contribution evaluated for the configuration $|\vec{k}_1|=|\vec{k}_2|=|\vec{k}_3|=|\vec{k}_4|$ with a varying $|\vec{k}_1+\vec{k}_2|$. The contribution from ${\cal O}_1$ and ${\cal O}_{2, \rm Bulk}$ match precisely while ${\cal O}_{2, \rm TD}$ does not contribute. {\it Right}. The real and imaginary parts of the ${\cal I}_{++}$ contribution. For the imaginary part, ${\cal O}_{2, \rm TD}$ does not contribute, and ${\cal O}_1$ and ${\cal O}_{2, \rm Bulk}$ match, as expected. For the real part, ${\cal O}_1$ matches with the sum of ${\cal O}_{2, \rm Bulk}$ and ${\cal O}_{2, \rm TD}$. The contribution from ${\cal O}_{2, \rm TD}$ does not depend upon either the momentum ratio (as seen here) or the mass parameter $\mu$ (as can be checked), indicating that it can be thought of as originating from a `local' contact interaction, such as the one shown schematically in the right panel of Fig.~\ref{fig_4pt}. For both the panels, we do not track the overall factors or momentum dependence.}
 \label{fig_compare}
\end{figure}

\paragraph{Contribution via ${\cal I}_{+-}$.}
 For ${\cal I}_{+-}$, the leading contribution involves two separate integrals, one for the time ordering, and the other for anti-time ordering. 
 Schematically they can be written as,
 \es{}{
 \langle {\cal O}_1\cdot \phi(\vec{k}_1) \phi(\vec{k}_2) \phi(\vec{k}_3) \phi(\vec{k}_4)\cdot  {\cal O}_1\rangle~{\rm and}~\langle {\cal O}_2\cdot \phi(\vec{k}_1) \phi(\vec{k}_2) \phi(\vec{k}_3) \phi(\vec{k}_4)\cdot  {\cal O}_2\rangle.
 }
 We evaluate these numerically and the comparison between ${\cal O}_1$ and ${\cal O}_2$ are shown in the left panel of Fig.~\ref{fig_compare}.
The four point function exhibits oscillations as function of the ratio of the momentum of the massive $\sigma$ particle, $|\vec{k}_1+\vec{k}_2|$ and the momentum of the inflaton $|\vec{k}_1|=k_1$. (For this evaluation we set $k_1 = k_2$.)
 This confirms a cosmological collider signature from this operator.
 We can also check explicitly that ${\cal O}_{2,\rm TD}$ does not contribute to ${\cal I}_{+-}$.
 To see this, note~\eqref{eq:O_2TD} implies ${\cal I}_{+-}$ has a contribution from,
 \es{}{
\propto {1\over \eta_0^2}\left[\partial_\eta(\phi^2)\sigma\right]_{\eta_0}.
 }
 However, this term decays as $\eta_0^{1/2}$ as $\eta_0\rightarrow 0$, and does not contribute to ${\cal I}_{+-}$.
 
\paragraph{Contribution via ${\cal I}_{++}$.}
Next we discuss ${\cal I}_{++}$ which involves nested time integrals, originating from the time-ordering operation.
Schematically, this has the form
\es{}{
\langle {\mathbb I}\cdot \phi(\vec{k}_1) \phi(\vec{k}_2) \phi(\vec{k}_3) \phi(\vec{k}_4)\cdot \int_{-\infty}^{\eta_0} \D \eta~{\cal O}_2(\eta) \int_{-\infty}^{\eta}\D \eta'~{\cal O}_2(\eta')\rangle,
}
where we have suppressed the spatial indices.
The operator ${\cal O}_1$ also contributes via a similar form.
From the above form, we note that ${\cal O}_{2,\rm TD}$ could contribute to correlation functions, since the inner integration contributes a non-vanishing integrand for the outer integral.
We evaluate this contribution from ${\cal O}_{2,\rm TD}$ numerically and indeed find it to be non-zero, as shown in the right panel of Fig.~\ref{fig_compare}.
There we also find the combined contributions from ${\cal O}_{2,\rm TD}$ and ${\cal O}_{2,\rm Bulk}$ match with ${\cal O}_{1}$.

Importantly, the contribution from ${\cal O}_{2,\rm TD}$ does not vary as a function of $|\vec{k}_1+\vec{k}_2|/k_1$, unlike the contribution from ${\cal O}_{2,\rm Bulk}$.
We also have checked that the contribution from ${\cal O}_{2,\rm TD}$ is independent of the mass of $\sigma$ for $m>\sqrt{2}H$.
 These facts indicate that the contribution from ${\cal O}_{2,\rm TD}$ could be thought of as coming from a contact operator that does not involve a $\sigma$ exchange.
 This can be understood in two different ways.
 
 First, instead of using the exact mode functions of $\sigma$, we can use their sub-horizon, high-energy limits.
 In detail, we can express $\sigma$ as,
 \es{}{
 \sigma(\eta,\vec{k}) = g_k(\eta) a_{\vec{k}}^\dagger  + \bar{g}_k(\eta)a_{-\vec{k}},
 }
 with 
 \es{eq:massive_mode}{
 g_k(\eta) & = +i \exp(-i\pi/4){\sqrt{\pi}\over 2}\exp(\pi \mu/2)(-\eta)^{3/2}H_{i\mu}^{(2)}(-k\eta),\\
 \bar{g}_k(\eta) & = -i \exp(+i\pi/4){\sqrt{\pi}\over 2}\exp(-\pi \mu/2)(-\eta)^{3/2}H_{i\mu}^{(1)}(-k\eta).
 }
 Using the high-energy limits of the Hankel functions (see, e.g.,~\cite{Chen:2022vzh} for explicit expressions), we can evaluate the contribution of ${\cal O}_{2,\rm TD}$.
 We find,
 \es{}{
 \langle \phi(\vec{k}_1) \phi(\vec{k}_2) \phi(\vec{k}_3) \phi(\vec{k}_4)\rangle \propto \left(-{13\over 256} + {k_1\over 16 |\vec{k}_1 + \vec{k}_2|}\right).
 }
The sub-horizon limit $|\vec{k}_1 + \vec{k}_2|(-\eta)\gg 1$, or equivalently $|\vec{k}_1 + \vec{k}_2|\gg k_1 \sim 1/(-\eta)$, then implies the leading contribution is given by the first term in the parenthesis above.
This then exactly reproduces the ${\cal O}_{2,\rm TD}$ contribution in the right panel of Fig.~\ref{fig_compare}.

Alternatively, we can construct a contact term that gives the same contribution as ${\cal O}_{2,\rm TD}$.
To that end, we first do a field redefinition:
\es{}{
\sigma \rightarrow \sigma + {c\over \Lambda} \phi^2.
}
Under this redefinition, the $\sigma$ kinetic term gives rise to,
\es{}{
{1\over 2}(\nabla_\mu\sigma)(\nabla^\mu\sigma)\rightarrow {1\over 2}(\nabla_\mu\sigma)(\nabla^\mu\sigma) + {c\over \Lambda}\nabla_\mu\sigma\cdot \nabla^\mu(\phi^2) + {c^2\over 2\Lambda^2}\nabla_\mu(\phi^2)\cdot \nabla^\mu(\phi^2).
}
For $c=1/2$, we reproduce the form of the bulk term ${\cal O}_{2,\rm Bulk}$.
This indicates that the action of ${\cal O}_{2,\rm TD}$ could be the same as the contact operator,
\es{}{
{\cal O}_{\rm contact}={1\over 8\Lambda^2}\nabla_\mu(\phi^2)\cdot\nabla^\mu(\phi^2).
}
The temporal component of the above operator gives the same contribution as ${\cal O}_{2,\rm TD}$ in Fig.~\ref{fig_compare}.

\section{IBP in Inflationary Spacetime: Examples} \label{app:ex}
In this appendix, we perform some explicit checks of IBP in inflationary spacetimes, taking into account the necessary boundary terms.
We first discuss massless fields (both free and interacting) and then massive fields.

\subsection{Massless Scalars}

\subsubsection{Free Theory}
A massless scalar in dS can be expanded in terms of the mode functions as,
\begin{align}
\label{modefunc}
    \phi(\eta, \vec{k}) = f_k(\eta)a_{\vec{k}}^\dagger + \bar{f}_k(\eta)a_{-\vec{k}},
\end{align}
with
\begin{align}
    f_k(\eta) = \frac{(1-ik\eta)e^{ik\eta}}{\sqrt{2k^3}},~\bar{f}_k(\eta) = \frac{(1+ik\eta)e^{-ik\eta}}{\sqrt{2k^3}}.
\end{align}
The free theory action, upon using the EOM $\square\phi = 0$, reduces to:
\begin{align}\label{eq:free_IBP}
    \int \D^4 x \sqrt{|g|} (\nabla_\mu\phi)^2 \stackrel{!}{=} \int \D^4 x \sqrt{|g|} \nabla_\mu(\phi\nabla^\mu \phi).
\end{align}
The $\stackrel{!}{=}$ indicates that we are interested in knowing whether both sides of the above equation would give the same correlation function or not.
To that end, we evaluate both sides on-shell, using the above mode functions of massless fields in dS.
The LHS gives
\begin{align}
    \int \frac{\D\eta \D^3 x}{\eta^2}\left[-(\partial_\eta\phi)^2+(\partial_i\phi)^2\right] = \int_{-\infty}^{\eta_0}\frac{\D\eta}{\eta^2}\left[-(k^2\eta)^2e^{2ik\eta}+k^2(1-ik\eta)^2e^{2ik\eta}\right]
\end{align}
After performing the integrals we arrive at
\begin{align}
    \int_{-\infty}^{\eta_0} \frac{\D\eta}{\eta^2}\left[-(\partial_\eta\phi)^2+(\partial_i\phi)^2\right] = -{k^2\over \eta_0}-ik^3.
\end{align}
The RHS of~\eqref{eq:free_IBP} is a boundary term.
We can use Stokes' theorem to write it as,
\es{}{
\int \D^4 x \sqrt{|g|} \nabla_\mu(\phi\nabla^\mu \phi) = \int \D^3 x \sqrt{|\gamma|}n_\mu (\phi\nabla^\mu \phi).
}
The induced metric on the boundary time slice is denoted by $\gamma$, with $\sqrt{|\gamma|} = 1/\eta_0^3$, and $n_\mu$ is a unit normal vector $n_\mu = (1/\eta_0, 0, 0, 0)$.
This can then be evaluated as $\eta_0\rightarrow 0$,
\begin{align}
    -\frac{1}{\eta_0^2}\phi\partial_\eta\phi\bigg\rvert_{\eta_0} = -\frac{k^2}{\eta_0}-ik^3.
\end{align}
This matches exactly with the LHS contribution.
Note the first term is naively divergent as $\eta_0\rightarrow 0$.
However it does not contribute to correlation functions, which in this context are the power spectrum.
This is because, to evaluate the power spectrum, we need to multiply the above by $(-i)$ (stemming from the $(-i)$ in $\exp(-i \int \D t {\mathbb H}))$ and sum with its conjugate.
The $1/\eta_0$ piece then would cancel.
The remaining term would contribute to the power spectrum as expected.

\subsubsection{Interactions}
As an example, we consider a massless scalar field $\phi$ in dS with an interaction term
\begin{align}
    \int \D^4x\sqrt{|g|}\nabla_\mu(\phi^2)\nabla^\mu\phi.
\end{align}
By IBP we can reduce this as,
\begin{align}\label{eq.ibp_check}
    \int \D^4x\sqrt{|g|}\nabla_\mu(\phi^2)\nabla^\mu\phi \stackrel{!}{=} \int \D^4x\sqrt{|g|}\nabla_\mu (\phi^2 \nabla^\mu \phi) = \int \D^3 x \sqrt{|\gamma|} n_\mu \phi^2\nabla^\mu\phi.
\end{align}
In the last relation, we have used Stokes' theorem. 
We have assumed $\phi$ vanishes at spatial infinity and dropped contribution from boundary terms at spatial infinity, keeping only the contribution from a late time slice at $\eta=\eta_0\rightarrow 0$, as above.
We can rewrite the boundary term as,
\begin{align}
    \int \D^3x \sqrt{|\gamma|} n_\mu \phi^2\nabla^\mu\phi = \int \D^3x \left(-\frac{1}{\eta_0^2}\right) \phi^2 \partial_\eta \phi\bigg\rvert_{\eta_0}.
\end{align}
We now compute the contact three-point interaction mediated by this operator and check the relation~\eqref{eq.ibp_check}.
\paragraph{LHS.}
Here we will not write $\int \D^3 x$ explicitly to keep the notation simple, and only track time integrals.
Then we can write the LHS of~\eqref{eq.ibp_check} as,
\begin{align}
    \int_{-\infty}^{\eta_0} \frac{\D\eta}{\eta^2}\left(-2\phi\partial_\eta\phi\partial_\eta\phi+2\phi\partial_i\phi\partial_i\phi\right)
\end{align}
Upon using the mode functions this becomes,
\begin{align}
    &\int_{-\infty}^{\eta_0} \frac{\D\eta}{\eta^2}\left(\left[(-2)(1-ik_1\eta)k_2^2\eta k_3^2\eta+{\rm perms.}\right]\right.\\
    &\left.+ \left[(2)(1-ik_1\eta)(1-ik_2\eta)(1-ik_3\eta)(-\vec{k}_2\cdot\vec{k}_3)+{\rm perms.}\right]\right)e^{ik_t\eta}
\end{align}
Here we have schematically included other momentum permutations and denoted $k_t = k_1+k_2+k_3$.
Using momentum conservation $\vec{k}_1 + \vec{k}_2 + \vec{k}_3 = 0$, the piece involving spatial derivatives can be simplified as,
\begin{align}
    \int_{-\infty}^{\eta_0}\frac{\D\eta}{\eta^2} \left[(1-ik_1\eta)(1-ik_2\eta)(1-ik_3\eta)(k_1^2+k_2^2+k_3^2)\right]e^{ik_t\eta}.
\end{align}
This can be evaluated as,
\begin{align}
    (+i)(k_1^2+k_2^2+k_3^2)\left[{i \over \eta_0}-k_t+\frac{k_1k_2+k_1k_3+k_2k_3}{k_t}+\frac{k_1k_2k_3}{k_t^2}\right].
\end{align}
The part involving time derivatives can be evaluated as,
\begin{align}
    (+i)(2k_2^2k_3^2)\left[\frac{1}{k_t}+\frac{k_1}{k_t^2}\right]+{\rm perms.}.
\end{align}
Summing the spatial and temporal contributions we have,
\begin{align}\label{eq:int_LHS}
   \int \frac{d\eta}{\eta^2}\left(-2\phi\partial_\eta\phi\partial_\eta\phi+2\phi\partial_i\phi\partial_i\phi\right) = -{(k_1^2+k_2^2+k_3^2)\over \eta_0}+(-i)(k_1^3 + k_2^3 + k_3^3).
\end{align}

\paragraph{RHS.}
The RHS of~\eqref{eq.ibp_check} can be evaluated as (we do not write $\int \D^3 x$ for brevity),
\begin{align}
    \left(-\frac{1}{\eta_0^2}\right)(1-ik_1\eta_0)(1-ik_2\eta_0)k_3^2\eta_0 e^{ik_t\eta_0}+\rm{perms.}.
\end{align}
Taking $\eta_0\rightarrow 0$ limit,
\begin{align}\label{eq.RHS_final}
 \left(-\frac{1}{\eta_0}\right)k_3^2(1+ik_3\eta_0)+{\rm{perms.}} = -\frac{(k_1^2+k_2^2+k_3^2)}{\eta_0}+(-i)(k_1^3+k_2^3+k_3^3).
\end{align}
This matches exactly with~\eqref{eq:int_LHS}.
Note the $1/\eta_0$ piece does not contribute to correlation functions for reasons identical to the free theory case discussed above.

\subsection{Massive Scalars}

\subsubsection{Free Theory}

We will start with a simple case, the kinetic term
\begin{align}
   \int  \D^4 x \sqrt{|g|} (\nabla_\mu \sigma)^2
\end{align}
for a massive scalar field $\sigma$. The mode functions for a massive scalar in dS are given in~\eqref{eq:massive_mode}.
We then check whether
\begin{align}
\label{eq:ibp_massive_scalar}
    \int \D^4 x \sqrt{|g|} (\nabla_\mu \sigma)^2 \stackrel{!}{=} \int \D^3x \sqrt{|\gamma|} n_\mu \sigma \nabla^\mu \sigma\bigg\rvert_{\eta_0} - \int \D^4 x \sqrt{|g|} \sigma \Box \sigma,
\end{align}
where $\Box\sigma \equiv \nabla_\mu\nabla^\mu\sigma$. 
To that end, instead of using the explicit forms of the Hankel functions, we will include them schematically and that will be sufficient for our purpose.
We will also check the equality~\eqref{eq:ibp_massive_scalar} by
analyzing the spatial and the temporal components separately.
The spatial part, again omitting $\int \D^3 x$ and tracking time integrals, is given by
\begin{align}
    \int \frac{\D\eta}{\eta^2} (\partial_i f_{k_1} \partial_i f_{k_2}) \stackrel{!}{=} -\int \frac{\D\eta}{\eta^2} f_{k_1} \partial_i^2 f_{k_2},
\end{align} 
Inserting the momentum factors, this becomes
\begin{align}
\label{eq:ibp_ms_spatial}
    \int \frac{\D\eta}{\eta^2} (\vec{k_1} \cdot \vec{k_2}) f_{k_1} f_{k_2} \stackrel{!}{=}  -\int \frac{\D\eta}{\eta^2} k_1^2 f_{k_1} f_{k_2}.
\end{align}
The two contributions are equal since momentum conservation forces $\vec{k}_1+\vec{k}_2=0$.
The temporal part is given by, 
\begin{align}
    -\int \frac{\D\eta}{\eta^2} \left( \partial_\eta f_{k_1} \partial_\eta f_{k_2}\right) \stackrel{!}{=}  \left(-\frac{1}{\eta_0^2}\right) f_{k_1} \partial_\eta f_{k_2}\bigg\rvert_{\eta_0} +  \int\frac{\D\eta}{\eta^2} f_{k_1} \nabla^2_\eta f_{k_2},
\end{align}
where $\nabla_\eta^2 = \partial_\eta^2 - \frac{2}{\eta} \partial_\eta$. 
Momentum conservation forces $k_1 = k_2$, and by doing an explicit temporal integration-by-parts, we can see the above equality indeed holds.

\subsubsection{Interacting Theory}
Next, we consider the interaction term
\begin{align}
    \int \D^4x\sqrt{|g|}\nabla_\mu(\phi^2)\nabla^\mu\sigma
\end{align}
We then follow the same procedure as used in the previous example. To check the validity of IBP for the case of massive scalars in dS, we consider whether

\begin{multline}
    \int \D^4 x \sqrt{|g|} \nabla_\mu(\phi^2)\nabla^\mu\sigma \stackrel{!}{=} \int \D^4 x \sqrt{|g|} \nabla_\mu (\phi^2 \nabla^\mu \sigma) - \int \D^4 x \sqrt{|g|} \phi^2 \square \sigma \\ = \int \D^3x \sqrt{|\gamma|} n_\mu \phi^2\nabla^\mu\sigma \bigg\rvert_{\eta_0} - \int \D^4 x \sqrt{|g|} \phi^2 \Box \sigma.
\end{multline}
Upon rewriting the boundary term, this condition becomes
\begin{align}
    \int \D^4 x \sqrt{|g|} \partial_\mu(\phi^2)\partial^\mu\sigma \stackrel{?}{=} \int \D^3x \left(-\frac{1}{\eta_0^2}\right) \phi^2 \partial_\eta \sigma\bigg\rvert_{\eta_0} - \int \D^4 x \sqrt{|g|} \phi^2 \Box \sigma.
\end{align}
We first compute the spatial part, again neglecting to write $\int \D^3 x$ for brevity. On the LHS, this is
\begin{multline}
    \int {\D\eta \over \eta^2} \partial_i (\phi^2) \partial_i \sigma = -\int \frac{\D\eta}{\eta^2} (\vec{k}_1 \cdot \vec{k}_3 + \vec{k}_2 \cdot \vec{k}_3) f_{k_1} f_{k_2} g_{k_3} = \int \frac{\D\eta}{\eta^2} k_3^2 f_{k_1} f_{k_2} g_{k_3},
\end{multline}
where we have used momentum conservation $\vec{k}_1+\vec{k}_3+\vec{k}_3=0$.
We have also considered a particular permutation of momenta, as this will be sufficient for our purpose.
On the RHS, we have
\es{eq:ibp_ms_int_spatial}
{- \int {\D\eta \over \eta^2} \phi^2 \partial_i^2 \sigma = \int \frac{\D\eta}{\eta^2} k_3^2 f_{k_1} f_{k_2} g_{k_3},} 
which matches with the LHS, as expected.

Now, we consider the temporal component. Starting again with the LHS and following a similar procedure as shown with the free theory, we have 
\es{}{
&- \int \frac{\D\eta}{\eta^2} \partial_\eta (f_{k_1} f_{k_2}) \partial_\eta g_{k_3} \\ &= \int \D\eta \partial_\eta \left(-\frac{1}{\eta^2} f_{k_1} f_{k_2} \partial_\eta g_{k_3} \right) + \int \D\eta f_{k_1} f_{k_2} \partial_\eta \left(\frac{1}{\eta^2} \partial_\eta g_{k_3} \right) \\ &= -\frac{1}{\eta_0^2} f_{k_1} f_{k_2} \partial_\eta g_{k_3}\bigg\rvert_{\eta_0} + \int \D\eta  f_{k_1} f_{k_2} \left(-\frac{2}{\eta^3} \partial_\eta g_{k_3} + \frac{1}{\eta^2} \partial_\eta^2 g_{k_3} \right)
}
We identify the first term as the boundary term and the second as the temporal component of $\square\sigma$.

In the above two examples --- a free theory and an interacting theory --- we have seen that IBP indeed holds for massive scalars in dS for contact diagrams.

\section{Dimension 5 Field Redefinition} 
The fact that ${\cal O}_{5,2}$ does not contribute to cosmological correlation functions can also be seen by performing a field redefinition.
In particular, we can redefine:
\es{eq:field_redef}{
A_\mu \rightarrow A_\mu - {\nabla_\mu \phi \over g_A \Lambda}.
}
Since we are not interested in late-time correlation functions of $A_\mu$, i.e., $A_\mu$ does not appear in the external lines, the above field redefinition does not modify inflaton correlators via contact diagrams.
It does have an effect on the other vertices.
While the kinetic terms for the inflaton and the gauge boson are unmodified by~\eqref{eq:field_redef}, 
\es{}{
|D_\mu {\cal H}|^2 &\rightarrow |D_\mu {\cal H}|^2 + {i \over \Lambda} \nabla^\mu \phi {\cal H}^\dagger \nabla_\mu {\cal H} - {i \over \Lambda} \nabla^\mu \phi {\cal H} \nabla_\mu {\cal H}^\dagger - {2 g \over \Lambda} A^\mu \nabla_\mu\phi {\cal H}^\dagger {\cal H} + {1 \over \Lambda^2} (\nabla_\mu\phi)^2 {\cal H}^\dagger {\cal H}\\
&= |D_\mu {\cal H}|^2 - {\cal O}_{5,2} + {1\over \Lambda^2} (\nabla_\mu\phi)^2 {\cal H}^\dagger {\cal H}.
}
Therefore, the field redefinition~\eqref{eq:field_redef} eliminates ${\cal O}_{5,2}$ and gives a correction to ${\cal O}_{6,1}$.

\end{appendix}
\bibliographystyle{utphys-modified}
\bibliography{references}
\end{document}